\documentclass[aps,twocolumn,preprintnumbers,nofootinbib,prd,superscriptaddress]{revtex4-1}

\usepackage[utf8]{inputenc}

\usepackage{graphicx,amssymb,amsmath,amsthm,amsfonts}
\usepackage{mathtools}
\usepackage[usenames,dvipsnames,svgnames,table]{xcolor}
\definecolor{darkred}{rgb}{0.5,0,0}
\usepackage{aas_macros}
\usepackage{bm}
\usepackage{dcolumn}
\usepackage{latexsym}
\usepackage{rotating}

\usepackage[normalem]{ulem}

\usepackage{enumerate}
\usepackage{tensor,multirow}
\usepackage{url}
\usepackage[linktocpage]{hyperref}

\newcommand{\LL}{\mathcal{L}}
\newcommand{\OO}{\mathcal{O}}

\newcommand{\gint}{\int_\Omega d^4x \, \sqrt{-g} \,}
\newcommand{\D}{\mathcal D}
\newcommand{\spatial}[1]{\prescript{(3)}{}{#1}}
\newcommand{\abs}[1]{\lvert #1 \rvert}
\newcommand{\tder}{\partial_{t}}
\newcommand{\rder}{\partial_{r}}

\usepackage{braket}

\def\be{\begin{equation}}
\def\ee{\end{equation}}
\newcommand{\beq}{\begin{eqnarray}}
\newcommand{\eeq}{\end{eqnarray}}

\def\ba{\begin{align}}
\def\ea{\end{align}}

\begin{document}

\title{Challenging the cosmic censorship in Einstein-Maxwell-scalar theory\\ with numerically simulated gedankenexperiments}

\author{Fabrizio Corelli}
\email{fabrizio.corelli@uniroma1.it}
\affiliation{Dipartimento di Fisica, ``Sapienza" Universit\`a di Roma \& Sezione INFN Roma1, Piazzale Aldo Moro 
5, 00185, Roma, Italy}

\author{Taishi Ikeda}
\affiliation{Dipartimento di Fisica, ``Sapienza" Universit\`a di Roma \& Sezione INFN Roma1, Piazzale Aldo Moro 
5, 00185, Roma, Italy}

\author{Paolo Pani}
\affiliation{Dipartimento di Fisica, ``Sapienza" Universit\`a di Roma \& Sezione INFN Roma1, Piazzale Aldo Moro 
5, 00185, Roma, Italy}

\begin{abstract}
We perform extensive nonlinear numerical simulations of the spherical collapse of (charged) wavepackets onto a charged black hole within Einstein-Maxwell theory and in Einstein-Maxwell-scalar theory featuring nonminimal couplings and a spontaneous scalarization mechanism. We confirm that black holes in full-fledged Einstein-Maxwell theory cannot be overcharged past extremality and no naked singularities form, in agreement with the cosmic censorship conjecture. We show that naked singularities do not form even in Einstein-Maxwell-scalar theory, although it is possible to form scalarized black holes with charge above the Reissner-Nordstr\"om bound. 
We argue that charge and mass extraction due to superradiance at fully nonlinear level is crucial to bound the charge-to-mass ratio of the final black hole below extremality.
We also discuss some ``descalarization'' mechanisms for scalarized black holes induced either by superradiance or by absorption of an opposite-charged wavepacket; in all cases the final state after descalarization is a subextremal Reissner-Nordstr\"om black hole.
\end{abstract}

\maketitle

\section{Introduction \& Executive Summary}
Thought experiments (also known as \emph{gedankenexperiments}~\cite{gedankenexperiments}) have always played a crucial role in the history of scientific discoveries. 
They have been of paramount importance in the development of new theories, in highlighting the crisis of old ones, or to elucidate particularly counterintuitive aspects of certain theories in a more accessible way. Limiting to physical sciences, notable examples are Newton's bucket, Schroedinger's cat, Einstein's elevator, Feynman's sprinkler, Dyson's sphere, etc.
The importance of gedankenexperiments relies on the fact that some deep consequences or internal inconsistencies of a theory can be explored by devising an ideal experiment regardless of the (im)possibility of its actual realization. In particular, the outcome of the experiment depends only on logic and on the given theoretical framework and is not affected by possible measurement errors or real-world noise.

Within General Relativity~(GR) a particularly relevant series of gedankenexperiments is devoted to test Penrose's cosmic censorship conjecture, according to which in four spacetime dimensions naked singularities (i.e. curvature singularities not covered by an event horizon) cannot form from typical regular initial data (see~\cite{Wald:1997wa} for an overview and a list of historical references). This conjecture has been put to the test by trying to overcharge/overspin a black hole~(BH) past extremality, by throwing test particles~\cite{1974AnPhy..82..548W,Hubeny:1998ga,Jacobson:2009kt,Saa:2011wq,Isoyama:2011ea,Natario:2016bay,Siahaan:2021bzc}, shells of matter~\cite{Hubeny:1998ga}, fluids~\cite{Aniceto:2015klq}, test fields~\cite{Duztas:2021kuj,Siahaan:2021bzc}, etc. Indeed, charged (resp., spinning) BHs in GR have a maximum amount of charge (resp., angular momentum) in units of their mass and above a critical value the Reissner-Nordstr\"om~(RN) (resp., Kerr) solution describes a naked singularity.

An ideal framework to perform gedankenexperiments are numerical simulations, since they allow to explore the dynamics of a full-fledged theory, without approximations that can ``contaminate'' the thought experiment.
For example, in the test-particle limit it is possible to overcharge/overspin a BH past extremality~\cite{Jacobson:2009kt}, but including backreaction and finite-size effects seems to rescue the cosmic censorship~\cite{Barausse:2010ka,Barausse:2011vx,Zimmerman:2012zu,Colleoni:2015afa,Sorce:2017dst,Colleoni:2015ena,Brito:2015oca,Vasquez:2021mjc,Sang:2021xqj}.
A particularly relevant question is whether the cosmic censorship is valid within GR in all cases, since Penrose's conjecture still lacks a formal proof. Another important question is whether the cosmic censorship is a prerogative of GR or whether it exists in some form also in other theories.
The latter point is particularly interesting given the fact that BHs in modified gravity are in general not described by the Kerr-Newman family. Furthermore, in recent years considerable attention has been put onto theories in which a spontaneous scalarization mechanism (originally devised for compact stars in scalar-tensor theories~\cite{Damour:1993hw,Damour:1996ke}) is at play also for BHs~\cite{Silva:2017uqg,Doneva:2017bvd,Antoniou:2017acq}. In these theories the Kerr-Newman solution coexists with other ``hairy BH'' solutions endowed with a scalar field (or with fields of other types~\cite{Ramazanoglu:2017xbl,Ramazanoglu:2019gbz}), which can be linearly stable and entropically favored over the standard GR BH solution, and can indeed be the endstate of a tachyonic instability affecting the latter.
Presently, little is known about the possibility of overcharging/overspinning a scalarized BH in these theories.

In this paper we perform gedankenexperiments in GR and in theories featuring a spontaneous scalarization mechanism, with the scope of challenging the cosmic censorship. We shall present numerical simulations that assume spherical symmetry but are otherwise exact and attempt to produce a naked singularity in various ways, especially by overcharging a BH with several wavepackets.

Our main results can be summarized as follows: 
\begin{itemize}
 \item[i)] We simulate the spherical collapse of an ingoing charged scalar field in an initially flat spacetime within Einstein-Maxwell theory, aiming at producing a BH that exceeds the RN bound. The final BH is always subextremal, confirming the results obtained in Ref.~\cite{Torres:2014fga}, which we extend to values of the final BH very close to extremality.
 \item[ii)] We performed extensive simulations trying to overcharge a RN BH within Einstein-Maxwell theory by throwing a charged scalar wavepacket. In this case a fraction of the wavepacket is repelled by the Coulomb interaction and the remaining part --~absorbed by the BH~-- is never sufficient to overcharge it past extremality, even when starting with nearly-extremal BHs. This again confirms the cosmic censorship in Einstein-Maxwell theory.
 \item[iii)] We repeated the same gedankenexperiment in an Einstein-Maxwell-scalar theory with spontaneous scalarization that allows for hairy charged BH solutions also with a charge above the RN limit. In this case we can form overcharged hairy BHs but in none of the simulations we observed the formation of naked singularities. We conclude that also in these theories the cosmic censorship is preserved, although the RN bound can be violated.
 \item[iv)] We unveil the crucial role played by BH charge and mass extraction due to superradiance at fully nonlinear level (see~\cite{Brito:2015oca} for an overview on superradiance) in preserving the cosmic censorship both in Einstein-Maxwell and in Einstein-Maxwell-scalar theory. As a by-product, we confirm and extend the results of~\cite{Baake:2016oku} for the superradiant amplification of charged wavepackets scattered off a charged BHs at the nonlinear level (see also Ref.~\cite{East:2013mfa} for a related study).
 \item[v)] For a fixed coupling constant, scalarized BHs in the nonminimally-coupled theories at hand exist only above a certain value of the charge-to-mass ratio~\cite{Herdeiro:2018wub,Fernandes:2019rez}. We show that these hairy BHs can ``descalarize'' either by absorbing opposite-charged wavepackets, or by a novel \emph{superradiantly-induced descalarization mechanism}. In all cases the endstate of descalarization is an ordinary RN BHs below the extremal limit, again confirming the cosmic censorship. 
\end{itemize}

We use geometric units with $G=c=4\pi\varepsilon_0=1$ and the Einstein summation convention throughout. In particular, Greek indices will run over the spacetime dimensions ($\mu,\nu,\cdots \in \{0, 1, 2, 3\}$), while Latin indices will run over the spatial dimensions ($i,j,\cdots \in \{1, 2, 3\}$). 
In Sec.~\ref{sec:setup} we present the field equations, our numerical scheme to evolve them, and discuss the initial and boundary conditions for the dynamical fields. The expert reader might wish to skip Sec.~\ref{sec:setup} and read directly Sec.~\ref{sec:results} where we present the results of various types of simulations.

\section{Setup} \label{sec:setup}

\subsection{Action of the theory and field equations}

We consider the Einstein-Maxwell-scalar model studied in Ref.~\cite{Herdeiro:2018wub}, minimally coupled to an additional (complex) charged scalar field:
\begin{multline}
	S = \frac{1}{16\pi}\gint \Bigl\{ R - 2 \bigl(\nabla_\mu \phi\bigr)\bigl(\nabla^\mu \phi\bigr) + \\ 
		- F[\phi] \, F_{\mu\nu}F^{\mu\nu} - 4 \bigl(\D_\mu \xi\bigr)\bigl(\D^\mu \xi\bigr)^* \Bigr\},
	\label{eq:action}
\end{multline}
where $\phi$ and $\xi$ are the real and the complex scalar fields respectively, $A_{\mu}$ is the vector field, $F[\phi]$ is the coupling function, $F_{\mu\nu} = \nabla_\mu A_\nu - \nabla_\nu A_\mu$ is the electromagnetic tensor, $\D_\mu = \nabla_\mu + i q A_\mu$ is the gauge covariant derivative for $U(1)$ symmetry, $q$ is the electric charge of the complex scalar field, and $\ast$ denotes the complex conjugate operation. $g_{\mu\nu}$ is the spacetime metric, and $R$ is the Ricci scalar.

The field equations that can be derived from \eqref{eq:action} are  
\begin{align}
	G_{\mu\nu} &= 8\pi \Bigl( T^\text{\tiny SF}_{\mu\nu} + T^\text{\tiny EM}_{\mu\nu} + T^{\xi}_{\mu\nu} \Bigr), \label{eq:field_grav} \\
	\nabla_\mu F^{\mu\nu} &= -F^{\mu\nu} \frac{1}{F[\phi]} \frac{\delta F[\phi]}{\delta \phi} \nabla_\mu \phi + \nonumber\\
		&+ \frac{i q}{F[\phi]} \Bigl[ \xi \bigl(\D^\nu \xi\bigr)^* - \xi^* \bigl(\D^\nu \xi \bigr) \Bigr], \label{eq:field_EM} \\
	\Box \phi &= \nabla_\mu \nabla^\mu \phi = \frac{1}{4} \frac{\delta F[\phi]}{\delta \phi} F_{\mu\nu}F^{\mu\nu}, \label{eq:real_field}\\
	\Box \xi &= - i q \bigl( \nabla_\mu A^\mu \bigr) \xi - 2 i q A^\mu \nabla_\mu \xi + q^2 A_\mu A^\mu \xi, \label{eq:complex_field} 
\end{align}
where $G_{\mu\nu} = R_{\mu\nu} - \frac{1}{2} R g_{\mu\nu}$ is the Einstein's tensor and
\begin{align}
	T^\text{\tiny SF}_{\mu\nu} &= \frac{1}{4\pi} \bigl( \nabla_\mu \phi \bigr) \bigl( \nabla_\nu \phi) - \frac{1}{8\pi} \bigl(\nabla_\alpha \phi\bigr)\bigl(\nabla^\alpha \phi\bigr) g_{\mu\nu}, \label{eq:TSF}\\
	T^\text{\tiny EM}_{\mu\nu} &= \biggl\{ \frac{1}{4\pi} F_{\mu\alpha} g^{\alpha\beta} F_{\nu\beta} - \frac{1}{16\pi}F_{\alpha\beta}F^{\alpha\beta} g_{\mu\nu} \biggr\} F[\phi], \label{eq:TEM} \\
	T^{\xi}_{\mu\nu} &= \frac{1}{4\pi} \Bigl[ \bigl(\D_\mu \xi\bigr)\bigl(\D_\nu \xi\bigr)^* + \bigl(\D_\mu \xi\bigr)^* \bigl(\D_\nu \xi\bigr) + \nonumber\\
			&- \bigl(\D_\alpha \xi\bigr)\bigl(\D^\alpha \xi\bigr)^* g_{\mu\nu} \Bigr]. \label{eq:Txi}
\end{align}

If $F[\phi]=1$, the model reduces to the well-studied Einstein-Maxwell theory minimally coupled to two (respectively neutral and charged) scalar fields.
In particular, the RN BH with $\phi=\xi=0$ is a stable solution of the theory with $F[\phi]=1$.
On the other hand, if $F[0]=1$ and $F''[0]>0$, the RN BH becomes unstable against spherical perturbations of the real scalar field,
and the scalarized charged BH might be favored~\cite{Herdeiro:2018wub}. 

It is worth mentioning that, when considering a spherically symmetric spacetime, the choice of a positive coupling function is a sufficent condition for the null energy condition to be satisfied. We report the proof of this statement in Appendix~\ref{app:NEC}. 
For the sake of generality for the moment we shall not assume any specific form of $F[\phi]$, but we shall require $F[0]=1$. In the result section we shall instead focus on the simplest model that gives rise to spontaneous scalarization, namely $F[\phi]=1-\lambda\phi^{2}$ with $\lambda<0$.
In this model, the null energy condition is satisfied.

While the real scalar field can trigger spontaneous scalarization of the BH, the complex scalar field is minimally coupled and is included in our setup only to change the charge of the BH. As such, stationary BH solutions in this theory have $\xi=0$.

\subsection{Evolution scheme}

For the time integration of the equations of motion we will use a generalization of the original Baumgarte-Shapiro-Shibata-Nakamura~(BSSN) formalism~\cite{Shibata:1995we,Baumgarte:1998te} in spherical symmetry~\cite{Brown:2009dd,Alcubierre:2011pkc}. The line element is given by
\begin{multline}
	ds^2 = (-\alpha^2 + \beta_r \beta^r) \, dt^2 + 2\beta_r\, dt\, dr + \\
	+ e^{4\chi (r, t)}\Bigl(a(r, t) \, dr^2 + b(r, t) \, r^2 \, d\Omega^2 \Bigr),
\end{multline}
where $\alpha$ is the lapse, $\vec \beta$ is the shift vector (which in spherical symmetry has only radial component), and $e^\chi$ is the conformal factor. The 3-metric of the spacelike hypersurfaces is $\gamma_{ij} = e^{4\chi}{\rm diag}(a,br^{2},br^{2}\sin^{2}\theta)$ and the lower radial component of $\vec \beta$ is given by $\beta_r = \gamma_{rr} \beta^r = e^{4\chi} a \beta^r$. 
Due to spherical symmetry, all functions depends on $(t,r)$ only. The metric functions $a$ and $b$ are initialized in such a way that the conformal metric $\hat{\gamma}_{ij} = e^{-4\chi} \gamma_{ij}$ is flat, and then in the evolution we considered the condition
\begin{equation}
	\tder \hat \gamma = (1 - \sigma) \Bigl(2\hat{\gamma} \hat{\nabla}_m \beta^m \Bigr),
	\label{eq:HatGammaEvolution}
\end{equation}
where $\hat{\gamma}$ is the derminant of $\hat{\gamma}_{ij}$, $\hat{\nabla}$ is the covariant derivative with respect to the conformal 3-metric, and $\sigma$ is a parameter that is set to $0$ for the so-called Eulerian evolution, and to $1$ for the Lagrangian evolution \cite{Brown:2009dd}. In the simulations described in this paper we used the latter. 

We also introduce the scalar and vector electromagnetic potentials
\begin{gather}
	\varphi = -n_\mu A^\mu, \\
	a^i = \tensor{\gamma}{^i_\mu} A^\mu,
\end{gather}
where $\tensor{\gamma}{^\mu_\nu}$ is the projector onto the foliation $\Sigma_t$,
and $n^{\mu}$ is the orthogonal vector of $\Sigma_{t}$.
The conjugate momenta of the real and complex scalar field are respectively defined as
\begin{gather}
	\Pi = n^\mu \nabla_\mu \phi, \\
	P = n^\mu \nabla_\mu \xi.
\end{gather}
With these definitions we can rewrite Eqs.~\eqref{eq:real_field} and \eqref{eq:complex_field} as two sets of first-order equations:
\begin{align}
	\tder \phi & = \beta^r \partial_r \phi + \alpha \Pi, \\
	\tder \Pi & = \beta^r \partial_r \Pi + \alpha \Pi K + \frac{(\partial_r \phi)(\partial_r \alpha)}{a \, e^{4\chi}} + \notag \\
		& + \frac{\alpha}{a \, e^{4\chi}} \Bigl[ \partial_r^2 \phi + \bigl(\partial_r \phi\bigr)\Bigl(\frac{2}{r} - \frac{\partial_r a}{2a} + \frac{\partial_r b}{b} + 2 \partial_r \chi\Bigr) \Bigr] + \notag \\
		&+ \frac{1}{2} \alpha a \, e^{4\chi} (E^r)^2 \frac{\delta F[\phi]}{\delta \phi},
	\label{eq:SFEOM}
\end{align}
for the real scalar field $\phi$, and
\begin{align}
	\tder \xi & = \beta^r \partial_r \xi + \alpha P, \\
	\tder P & = \beta^r \partial_r P + \alpha P K + \frac{(\partial_r \xi)(\partial_r \alpha)}{a \, e^{4\chi}} + \notag \\
		& + \frac{\alpha}{a \, e^{4\chi}} \Bigl[ \partial_r^2 \xi + \bigl(\partial_r \xi\bigr)\Bigl(\frac{2}{r} - \frac{\partial_r a}{2a} + \frac{\partial_r b}{b} + 2 \partial_r \chi\Bigr) \Bigr] + \notag \\
		& + 2 i q \alpha \Bigl( \varphi P + \frac{a_r \rder \xi}{a e^{4 \chi}} \Bigr) - q^2 \alpha \Bigl( \frac{(a_r)^2}{a e^{4 \chi}} - \varphi^2 \Bigr) \xi\,,
	\label{eq:CSFEOM}
\end{align}
for the complex scalar field $\xi$. Here $K$ is the trace of the extrinsic curvature $K_{ij}$.

Due to spherical symmetry the magnetic field vanishes and the only nonvanishing component of the electric field and of the vector electromagnetic potential is the radial one. Fixing the gauge with the Lorenz condition $\nabla_\mu A^\mu = 0$, we can write the equations of motion for the electromagnetic field as
\begin{align}
	D_i E^i &= -E^r \bigl(\partial_r \phi\bigr) \frac{1}{F[\phi]} \frac{\delta F[\phi]}{\delta \phi} + \notag \\
		&+ \frac{iq}{F[\phi]} \Bigl[ - \xi^* P + \xi P^* + 2 i q \varphi \abs{\xi}^2 \Bigr] \label{eq:E_constraint}\,,\\
	\partial_t E^r &= \alpha K E^r + \beta^r \partial_r E^r - E^r \partial_r \beta^r + \notag \\
		&- \alpha \Pi E^r \frac{1}{F[\phi]} \frac{\delta F[\phi]}{\delta \phi} + \frac{\alpha}{F[\phi]} \frac{2q^2}{a e^{4 \chi}} \abs{\xi}^2 a_r + \notag\\
		&+ \frac{\alpha}{F[\phi]} \frac{iq}{a e^{4 \chi}} \Bigl[ \xi \bigl(\rder \xi \bigr)^* - \xi^* \bigl(\rder \xi \bigr) \Bigr]\label{eq:E_evolution}\,, \\
	\tder a_r &= \beta^r \rder a_r + a_r \rder \beta^r - \rder (\alpha \varphi) - \alpha a e^{4 \chi} E^r \label{eq:ar_evolution}\,,\\
	\tder \varphi &= \beta^r \rder \varphi + \alpha \varphi K - \frac{(\rder \alpha ) a_r}{a e^{4 \chi}} + \notag\\ 
	 	&- \frac{\alpha}{a e^{4 \chi}} \biggl[ \rder a_r + a_r \biggl(\frac{2}{r} - \frac{\rder a}{2 a} + \frac{\rder b}{b} + 2 \rder \chi \biggl) \biggr] \label{eq:varphi_evolution}\,,
\end{align}
where $D_i$ is the covariant derivative with respect to the 3-metric $\gamma_{ij}$.
Equations~\eqref{eq:E_constraint} and \eqref{eq:E_evolution} have been obtained by projecting the field equation for the electromagnetic field \eqref{eq:field_EM} onto $n^\mu$ and onto $\Sigma_t$, respectively. The evolution equation for $a_r$ has been obtained from the definition of the electric field $E^\nu = - n_\mu F^{\mu\nu}$, while Eq.~\eqref{eq:varphi_evolution} has been derived from the Lorenz gauge condition~\cite{Alcubierre:2009ij, Torres:2014fga}. 

For the gravitational field we used the equations of the generalized BSSN formalism in spherical symmetry~\cite{Brown:2009dd,Alcubierre:2011pkc}. 
Introducing the traceless conformal extrinsic curvature $\hat{A}_{ij} = e^{-4\chi}(K_{ij}-\frac{1}{3}K\gamma_{ij})$, we define $A_a = \tensor{\hat{A}}{^r_r}$ and $A_b = \tensor{\hat{A}}{^\theta_\theta}$. These two variables are not independent, since $\hat{A}_{ij}$ is traceless and $A_a + 2A_b = 0$, therefore we only evolved $A_a$. We also introduce the BSSN variable 
\begin{equation}
	\hat{\Delta}^i = \hat{\gamma}^{mn} (\hat{\Gamma}^i_{mn} - \mathring{\Gamma}^i_{mn}),
	\label{eq:HatDelta}
\end{equation}
where $\hat{\Gamma}^i_{mn}$ and $\mathring{\Gamma}^i_{mn}$ are the Christoffel symbols of the conformal and the flat metrics, respectively. 

Having fixed the notation, we can now write the evolution equations for the gravitational sector as
\begin{align}
	\tder \chi &= \beta^r \rder \chi - \frac{1}{6} \alpha K  + \frac{\sigma}{6} \hat \nabla_m \beta^m, \label{eq:chi} \\
	\tder a &= \beta^r \rder a + 2 a \rder \beta^r - 2 \alpha a A_a - \frac{2}{3} \sigma a \hat \nabla_m \beta^m, \label{eq:a} \\
	\tder b &= \beta^r \rder b + 2 b \frac{\beta^r}{r} - 2 \alpha b A_b - \frac{2}{3} \sigma b \hat \nabla_m \beta^m, \label{eq:b} \\
	\tder K &= \beta^r \rder K - D^2 \alpha + \alpha \Bigl( A_a^2 + 2 A_b^2 + \frac{1}{3} K^2 \Bigr) + \notag\\
		&+ 4 \pi \alpha (S_a + 2 S_b + \mathcal{E}), \label{eq:K} \\
	\tder A_a &= \beta^r \rder A_a + \alpha K A_a -  \Bigl(D^r D_r \alpha - \frac{1}{3} D^2 \alpha \Bigr) + \notag\\
		& + \alpha (\tensor{R}{^r_r} - \frac{1}{3} R) - \frac{16 \pi \alpha}{3} \Bigl(S_a - S_b \Bigr), \label{eq:Aa} \\
	\tder \hat \Delta^r &= \beta^r \rder \hat \Delta^r - \hat \Delta^r \rder \beta^r + \frac{2}{b} \rder \biggl(\frac{\beta^r}{r} \biggr) + 2 \alpha A_a \hat \Delta^r + \notag\\
		& -2\alpha \Bigl( A_a - A_b \Bigr) \frac{2}{br} - \frac{2}{a} \Bigl( A_a \rder \alpha + \alpha \rder A_a \Bigr) + \notag\\
		& + \frac{1}{a} \partial^2_r \beta^r +\frac{\sigma}{3} \biggl[ \frac{1}{a} \rder \hat \nabla_m \beta^m + 2 \hat \Delta^r \hat \nabla_m \beta^m \biggr] + \notag\\
		& + \frac{2\alpha}{a} \biggl[ \rder A_a + \Bigl(A_a - A_b \Bigr)\biggl(\frac{\rder b}{b} + \frac{2}{r} \biggr) + \notag \\
		&+ 6 A_a \rder \chi - \frac{2}{3} \rder K - 8 \pi j_r \biggl] \label{eq:Delta},
\end{align}
where $R_{ij}$ and $R$ are respectively the Ricci tensor and the scalar curvature of the 3-metric $\gamma_{ij}$, and the constraint equations read
\begin{align}
	H &= R + \frac{2}{3} K^2 - (A_a^2 + 2 A_b^2) - 16 \pi \rho = 0, \label{const:H}\\
	M &= \rder A_a + \Bigl(A_a - A_b \Bigr)\biggl(\frac{\rder b}{b} + \frac{2}{r} \biggr) + \notag \\
	  &+ 6 A_a \rder \chi - \frac{2}{3} \rder K - 8 \pi j_r = 0. \label{const:M} 
\end{align}
The source terms can be divided into three contributions:
\begin{itemize}
	\item from the electromagnetic field we have
		\begin{align}
			\mathcal{E}^\text{\tiny EM} &= n^\mu n^\nu T^\text{\tiny EM}_{\mu\nu} = \frac{1}{8\pi} a \, e^{4 \chi} (E^r)^2 F[\phi], \\
			S^\text{\tiny EM}_a &= \tensor{\Bigl( \spatial{T}^\text{\tiny EM} \Bigl)}{^r_r} = -\frac{1}{8\pi} a \, e^{4 \chi} (E^r)^2 F[\phi], \\
			S^\text{\tiny EM}_b &= \tensor{\Bigl( \spatial{T}^\text{\tiny EM} \Bigl)}{^\theta_\theta} = \frac{1}{8\pi} a \, e^{4 \chi} (E^r)^2 F[\phi], 
			\label{eq:EMSources} 
		\end{align}
	\item from the real scalar field we have
		\begin{align}
			\mathcal{E}^\text{\tiny SF} &= n^\mu n^\nu T^\text{\tiny SF}_{\mu\nu} = \frac{1}{8\pi} \biggl( \Pi^2 + \frac{(\partial_r \phi)^2}{a \, e^{4\chi}} \biggr),\\
			j^\text{\tiny SF}_r &= - \tensor{\gamma}{^\mu_r} n^\nu T^\text{\tiny SF}_{\mu\nu} = -\frac{1}{4\pi} \Pi \partial_r \phi, \\
			S^\text{\tiny SF}_a &= \tensor{\Bigl( \spatial{T}^\text{\tiny SF} \Bigl)}{^r_r} = \frac{1}{8\pi} \biggl( \Pi^2 + \frac{(\partial_r \phi)^2}{a \, e^{4\chi}} \biggr),\\
			S^\text{\tiny SF}_b &= \tensor{\Bigl( \spatial{T}^\text{\tiny SF} \Bigl)}{^\theta_\theta} = \frac{1}{8\pi} \biggl( \Pi^2 - \frac{(\partial_r \phi)^2}{a \, e^{4\chi}} \biggr),
			\label{eq:RSFSources} 
		\end{align}
	\item and from the complex scalar field
		\begin{align}
			\mathcal{E}^{\xi} &= n^\mu n^\nu T^{\xi}_{\mu\nu} = \frac{1}{4\pi} \biggl( \abs{\tilde P}^2 + \frac{\abs{\tilde \Psi}^2}{a \, e^{4\chi}} \biggr),\\
			j^{\xi}_r &= - \tensor{\gamma}{^\mu_r} n^\nu T^{\xi}_{\mu\nu} = -\frac{1}{4\pi} \Bigl(\tilde \Psi \tilde P^* + \tilde P \tilde \Psi^*  \Bigr), \\
			S^{\xi}_a &= \tensor{\Bigl( \spatial{T}^{\xi} \Bigl)}{^r_r} = \frac{1}{4\pi} \biggl( \abs{\tilde P}^2 + \frac{\abs{\tilde \Psi}^2}{a \, e^{4\chi}} \biggr),\\
			S^{\xi}_b &= \tensor{\Bigl( \spatial{T}^{\xi} \Bigl)}{^\theta_\theta} = \frac{1}{4\pi} \biggl( \abs{\tilde P}^2 - \frac{\abs{\tilde \Psi}^2}{a \, e^{4\chi}} \biggr),
			\label{eq:CSFSources} 
		\end{align}
    where we have defined the terms
		\begin{align}
			\tilde P &= n^\mu \D_\mu \xi = n^\mu \nabla_\mu \xi + i q n^\mu A_\mu \xi = P - i q \varphi \xi,\\
			\tilde \Psi &= \tensor{\gamma}{^\mu_r}\D_\mu \xi = \tensor{\gamma}{^\mu_r}\nabla_\mu \xi + i q \tensor{\gamma}{^\mu_r} A_\mu \xi = \rder \xi + i q a_r \xi.
		\end{align}
\end{itemize}
Note that, for practical reasons, in our code we evolved the variable $ e^{-2\chi}$ instead of $\chi$.

For the evolution of the lapse function we use the nonadvective 1+log slicing condition~\cite{Bona:1994dr}
\begin{equation}
	\tder \alpha = - 2 \alpha K,
\end{equation}
while for the shift we use the Gamma-driver condition~\cite{Alcubierre:2002kk, Alcubierre:2011pkc}; namely we define a new variable $B^r$ such that
\begin{align}
	\tder B^r &= \frac{3}{4} \tder \hat \Delta^r, \\
	\tder \beta^r &= B^r.
\end{align}

\subsection{Electric charge in Einstein-Maxwell-scalar theory} \label{sec:ElectricCharge}

Due to the nonminimal coupling, in this theory it is possible to define the electric charge in two different ways. The equation for the electromagnetic field can in fact be written as $\nabla_\mu F^{\mu\nu} = -4 \pi J_\text{\tiny EM}^\nu$, where
\begin{align}
	J_\text{\tiny EM}^\nu &= \frac{1}{4\pi} \biggl\{\frac{1}{F[\phi]} \frac{\delta F[\phi]}{\delta \phi} \bigl(\nabla_\mu \phi \bigr) F^{\mu\nu} + \notag \\
		   &- \frac{i q}{F[\phi]} \Bigl[ \xi \bigl(\D^\nu \xi \bigr)^* - \xi^* \bigl(\D^\nu \xi \bigr) \Bigr] \biggr\},
	\label{eq:EmCurrent}
\end{align}
but also as $\nabla_\mu \Bigl(F[\phi] F^{\mu\nu}\Bigr) = - 4 \pi \tilde J_\text{\tiny EM}^\nu$, where
\begin{equation}
	\tilde J_\text{\tiny EM}^\nu = - \frac{i q}{4\pi} \Bigl[ \xi \bigl(\D^\nu \xi \bigr)^* - \xi^* \bigl(\D^\nu \xi \bigr) \Bigr].
	\label{eq:EmTildeCurrent}
\end{equation}
Both these two currents are conserved, namely $\nabla_\mu J_\text{\tiny EM}^\mu = 0 = \nabla_\mu \tilde J_\text{\tiny EM}^\mu$, and allow to define the electric charge in two ways:
\begin{align}
	Q &= \frac{1}{4\pi} \int_V dV D_i E^i = \int_V dV \rho \label{eq:Qdef},\\
	\tilde Q &= \frac{1}{4\pi} \int_V dV D_i \bigl( F[\phi] E^i \bigr) = \int_V dV \tilde \rho \label{eq:TildeQdef},
\end{align}
where the two charge densities are
\begin{align}
	\rho = -n_\mu J_\text{\tiny EM}^\mu &= \frac{1}{4\pi} \biggl\{ -E^r \bigl(\partial_r \phi\bigr) \frac{1}{F[\phi]} \frac{\delta F[\phi]}{\delta \phi} + \notag \\
	     &+ \frac{iq}{F[\phi]} \Bigl[ - \xi^* P + \xi P^* + 2 i q \varphi \abs{\xi}^2\Bigr] \biggr\} \label{eq:rhodef}, \\
	\tilde \rho = -n_\mu \tilde J_\text{\tiny EM}^\mu &= \frac{iq}{4\pi} \Bigl[ - \xi^* P + \xi P^* + 2 i q \varphi \abs{\xi}^2\Bigr] \label{eq:Tilderhodef}.
\end{align}
As it can be seen from the above equations, while the charge $Q$ includes the contribution of the real scalar field, $\tilde Q$ accounts only for the charge carried by the complex field $\xi$. In Einstein-Maxwell theory ($F[\phi] = 1$) or when the scalar field vanishes ($F[\phi = 0] = 1$), the two charges coincide, as expected.  

For a spherically symmetric spacetime, following \cite{Torres:2014fga}, we can define the electric charge enclosed in the 2-sphere $S_r$ of radius $r$ in two ways:
\begin{align}
	Q(r) &= \int_{S_r} dV \rho = \frac{1}{4\pi} \int_{S_r} dV D_i E^i = \notag \\
	     &= \frac{1}{4\pi} \int_{\partial S_r} dS \, s_i E^i = \sqrt{a} b e^{6\chi} r^2 E^r, \notag \\ 
	\tilde{Q}(r) &= \int_{S_r} dV \tilde \rho = \frac{1}{4\pi} \int_{S_r} dV D_i \bigl( F[\phi] E^i \bigr) = \notag \\
		&= \frac{1}{4\pi} \int_{\partial S_r} dS \, s_i E^i F[\phi] = F[\phi] \sqrt{a} b e^{6\chi} r^2 E^r,
	\label{eq:Qr}
\end{align}
where $s^i$ is the outward pointing unit vector normal to $\partial S_r$. Note that, although we only made the radial dependence explicit, the above quantities can generically depend also on the time coordinate.

We can see that the electric field can be written as 
\begin{equation}
	E^r = \frac{Q(r)}{b\sqrt{a} e^{6\chi} r^2} = \frac{\tilde{Q}(r)}{F[\phi] b\sqrt{a} e^{6\chi} r^2}.
	\label{eq:ErEMS}
\end{equation}
and that the two definitions of charge can be related by
\begin{equation}
	Q(r) = \frac{\tilde Q(r)}{F[\phi]}.
	\label{eq:ChargesRealtion}
\end{equation}
For a spherically symmetric BH spacetime with a vanishing complex scalar field, $\tilde Q$ is homogeneous outside the horizon, while $Q$ is in general a radial function. For a scalarized configuration there is a nonvanishing charge density $\rho$ outside the BH and the total charge of the system does not coincide with the charge enclosed in the horizon. However, the two charges coincide at infinity since we shall always assume asymptotic flatness and hence $\phi\to0$ and $F[\phi] \to1$ and $r\to+\infty$.

\subsection{Numerical integration scheme}

In our framework the equations of motion are regular at the origin, but contain terms that go as $\frac{1}{r}$ and $\frac{1}{r^2}$, that can cause instabilities in the numerical integration. To handle these terms we used the second-order Partially Implicitly Runge-Kutta (PIRK) method~\cite{Montero:2012yr, Cordero-Carrion:2012qac}, which does not require the implementation of an explicit regularization procedure at the origin. This allows us to integrate the equations that contain unstable terms with a partially implicit method, while the other equations can be integrated with an explicit method. The details of this implementation can be found in Appendix~\ref{app:PIRK}. 

For the numerical radial derivatives we used the fourth-order accurate centered finite differences method, except for the advection terms (which are of the form $\beta^r \rder $) for which we used the upwind scheme. In order to avoid the appearance of high-frequencies instabilities in the evolution, we added to all the equations a Kreiss-Oliger dissipation term, that we evolved explicitly; in this term the fourth derivative has been computed with second-order accuracy.
In Appendix~\ref{app:convergence} we show the numerical convergence of our code.

\subsection {Initial conditions}

Since our purpose is to study the collapse of a charged scalar field and the possibility of forming overcharged BH solutions, we choose an initial profile for $\xi$ that carries a nonvanishing amount of electric charge and propagates toward the horizon:
\begin{align}
	\xi(r, t = 0) &= B_0 e^{-\frac{1}{2} \sigma_\xi^2 (r-r_{0, \xi})^2+i k_0 (r-r_{0, \xi})}, \notag \\
	P(r, t = 0) &= i B_0 e^{-\frac{1}{2} \sigma_\xi^2 (r-r_{0, \xi})^2+i k_0 (r-r_{0, \xi})} \times \notag \\
		    & \times \left(k_0+i \sigma_\xi^2 (r-r_0)\right),
	\label{eq:InitialXi}
\end{align}
where $B_{0}$, $\sigma_{\xi}^{-1}$, $k_{0}$, and $r_{0,\xi}$ are respectively the amplitude, width, frequency, and position of the initial profile of the complex scalar field.

We choose a vanishing initial shift and a flat conformal 3-metric. We set to zero the auxiliary variable $B^r$ and the radial component of the traceless extrinsic curvature $A_a$, while we initialized $\hat \Delta^r$ using its definition in Eq.~\eqref{eq:HatDelta}, which in spherical symmetry reduces to~\cite{Alcubierre:2011pkc}
\begin{equation}
	\hat \Delta^r = \frac{1}{a} \biggl[ \frac{\rder a}{2a}  - \frac{\rder b}{b} - \frac{2}{r} \biggl( 1 - \frac{a}{b} \biggr) \biggr].
	\label{eq:DeltaDefinition}
\end{equation}

To find the initial profile of the electric field, the trace of the extrinsic curvature, and the conformal factor we solved Eq.~\eqref{eq:E_constraint} together with the Hamiltonian and momentum constraints. 
We also initialize the electromagnetic potentials to a configuration such that both $\varphi$ and $a_r$ do not evolve in a region sufficiently far from the horizon as long as the signals do not reach the outer boundary. To achieve this we set $a_r = 0$ at $t = 0$, and we determined the profile of $\varphi$ by solving the equation $\tder a_r = 0$ which, using Eq.~\eqref{eq:ar_evolution}, reduces to 
$\rder (\alpha \varphi) = - \alpha a e^{4 \chi} E^r$. 

The system of equations that we solved at $t = 0$ for $E^r$, $K$, $\varphi$, and $\psi := e^{\chi}$ reads
\begin{align}
	\rder^2 \psi &= \frac{1}{48 r^2 a b^2}  \biggl\{2 a^2 b \psi  \bigl[r^2 b \psi ^4 \bigl(-48 \pi  \mathcal E +2 K^2\bigr)+6\bigr] + \notag \\
		     &+ 6 r (\rder a) b \bigl[r (\rder b) \psi +2 b \bigl(2 r (\rder \psi)+\psi \bigr)\bigr] + \notag \\
		     & -3 a \Bigl[-r^2 (\rder b)^2 \psi +4 b^2 \left(8 r (\rder \psi)+\psi \right) + \notag \\
		     &+4 r b \Bigl(4 r (\rder b) (\rder \psi)+\left(3 \rder b+r \rder^2 b\right) \psi \Bigr)\Bigr]\biggr\}, \label{eq:InitialConstraints psi}\\
	\rder E^r &= - \biggl(\frac{\rder a}{2a} + \frac{\rder b}{b} + 6 \frac{\rder \psi}{\psi} + \frac{2}{r} \biggr) E^r+ \notag \\
		  &+ 2 q \frac{\xi_R P_I - \xi_I P_R}{F[\phi]} - 2 q^2 \varphi \frac{\abs{\xi}^2}{F[\phi]} + \notag \\
		  &- E^r (\rder \phi)\frac{1}{F[\phi]} \frac{\delta F[\phi]}{\delta \phi}, \\ 
	\rder K &= 6 \bigl[P_R \rder \xi_R + P_I \rder \xi_I - q \varphi (\xi_R \rder \xi_I - \xi_I \rder \xi_R)\bigr], \\ 
	\rder \varphi &= -\frac{\rder \alpha}{\alpha} \varphi - a \psi^4 E^r,
	\label{eq:InitialConstraints}
\end{align}
where the subscripts ${X}_R$ and ${X}_I$  denote the real and imaginary part of a complex variable $X$, respectively.

\subsection{Boundary conditions}

Thanks to the PIRK integration method at the origin we only impose the parity condition related to the spherical symmetry. Therefore we shifted the numerical grid in such a way that the origin is placed in the middle of a grid step, and the first grid point is at $r_1 = \frac{\Delta r}{2}$, where $\Delta r$ is the grid step. To compute the numerical derivatives at $r_1$ and $r_2$ we added two ghost grid points at $r_{-1} = - \frac{\Delta r}{2}$ and $r_{-2} = - \frac{3 \Delta r}{2}$ in which the variables are not evolved but are set at each timestep to values that satisfy the parity conditions. In particular $\beta^r$, $E^r$, $B^r$, and $\hat \Delta^r$ have odd parity at the origin while all the other variables have even parity.

At the outer boundary we added four ghost zones which are used to compute the fourth-order accurate upwind derivatives. In these zones the variables are not evolved and they remain constant. This can be done since we consider an initial profile of $\varphi$ such that the electromagnetic potentials do not evolve at the outer boundary as long as the signals coming from the horizon region are sufficiently far from the outer boundary, and we consider a domain large enough that outward-moving components of the initial field profiles do not reach the outer boundary during the time of integration.

\section{Results} \label{sec:results}

\subsection{Collapse of the charged field in a flat background in Einstein-Maxwell theory}

We start by neglecting the real field ($\phi=0$, $F[0]=1$) and study the collapse of the complex scalar field in flat spacetime in Einstein-Maxwell theory,
in order to explore the RN BH formation and the robustness of the cosmic censorship hypothesis in the standard case.
This problem was studied in~\cite{Torres:2014fga} using momentarily static charged wavepackets as the initial data.
In that case it was possible to form a RN BH with final charge-to-mass ratio as large as $Q/M\sim 0.6$, therefore still far from extremality.
In our simulation, we start from an ingoing charged wavepacket, so we expect that we could form a BH with higher charge, which is a more stringent test of the cosmic censorship.

\subsubsection{Initial setup}
We define an arbitrary mass scale $M$ to normalize all dimensionful quantities. We chose the parameters in Eq.~\eqref{eq:InitialXi} in such a way that the initial profile of $\xi$ is narrow enough to obtain final configurations in which the (possibly formed) final BH is close to extremality. In particular we set 
\begin{align}
 &B_0 = 0.012\,, \quad  k_0 M = 5\,, \notag \\
 &\sigma_\xi^2 M^2 = 2.5 \,,\quad r_{0, \xi}/M = 5\,.
\end{align}

\begin{figure}
	\centering
	\includegraphics[width = \columnwidth]{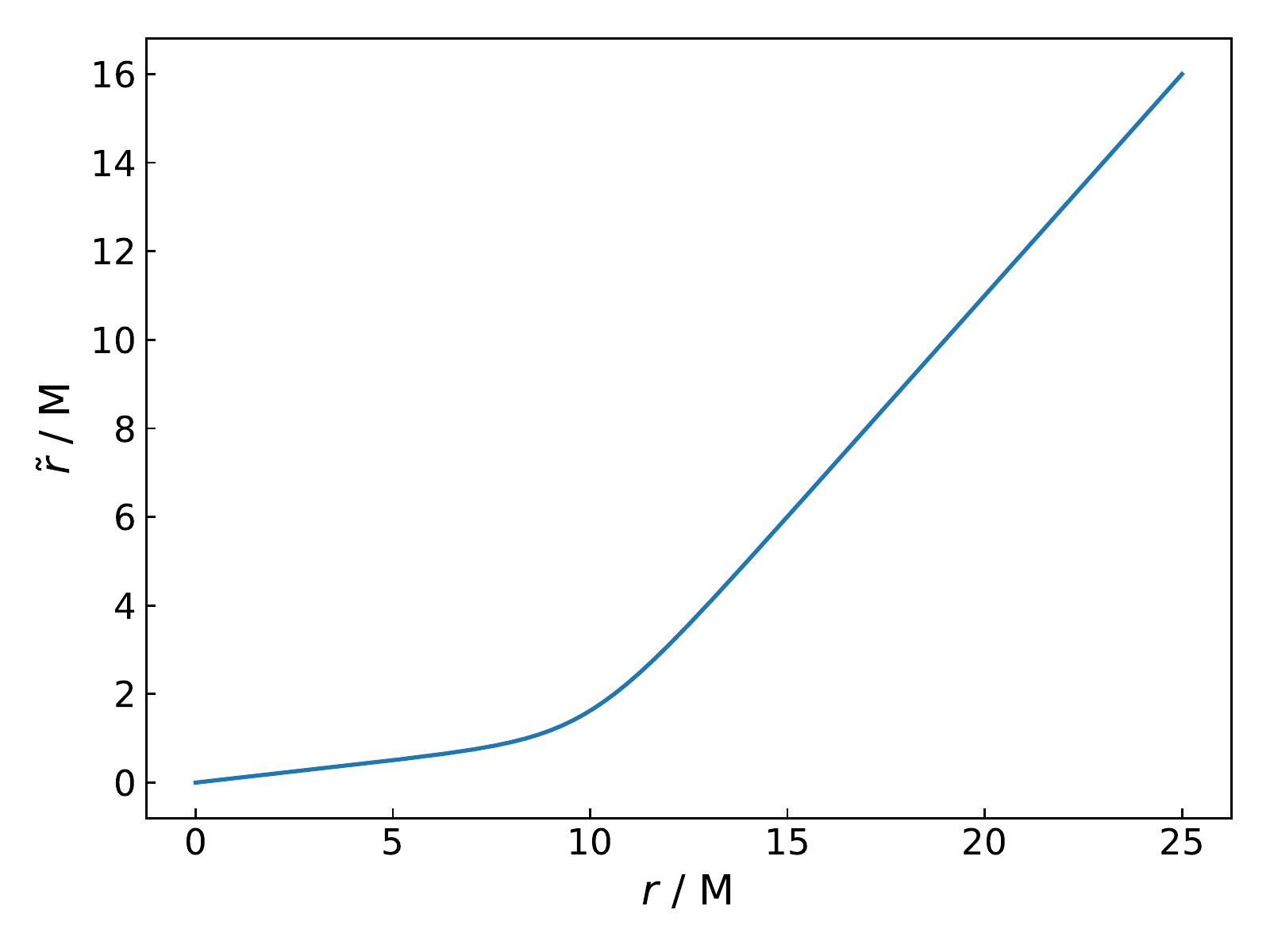}
	\caption{Coordinate transformation for the nonuniform grid step. Near the origin small regions in the (original) $\tilde r$ domain are mapped to large regions of the (new) $r$ domain. Sufficiently far from the origin the two coordinates differ only by a constant.
	}
	\label{fig:rtilde_plot}
\end{figure}

For this initial configuration the simulation is computationally demanding: high resolution and a low Courant–Friedrichs–Lewy~(CFL) factor are required. Therefore in order to obtain higher accuracy without increasing excessively the computational cost, we use a nonuniform grid step by performing the following transformation on the radial coordinate: 
\begin{equation}
	\begin{cases}
		\tilde r &= C(r) =  r + \frac{1 - \eta}{\Delta} \ln \Bigl( \frac{1 + e^{-\Delta (r - R_1)}}{1 + e^{\Delta R_1}} \Bigr) \\ 
		\frac{\partial \tilde r}{\partial r} &= C'(r) = \eta + \frac{1 - \eta}{1 + e^{-\Delta (r - R_1)}}
	\end{cases}\,,
\end{equation}
where we renamed the new radial coordinate as $r$ and the old one as $\tilde r$. 
In the above equation, $R_{1}$ and $\Delta$ are the typical radius and typical width of the buffer zone between the area around the origin that requires the higher numerical resolution and the asymptotic region, whereas $\eta$ characterizes the relative scale of the resolution.
The parameters are set to $\eta = 0.1$, $\Delta= 1/M$ and $R_1 = 10M$. The behavior of $\tilde r$ vs $r$ is shown in Fig.~\ref{fig:rtilde_plot}, where it can be seen that a small region around the center in the old coordinate $\tilde r$ is mapped to a larger region in the new coordinates. In this way the horizon of a final BH which is close to extremality is placed at a higher value of $r$ allowing for higher accuracy with a larger grid step. On the other hand $C'(r) \sim 1$ for $r \gg R_1$, and the two radial coordinates differ only by a constant near the outer boundary.
After this change of coordinates the metric functions $a$ and $b$ of the flat spacetime are:
\begin{align}
	a(r, t = 0) &= C'(r)^2, \\
	b(r, t = 0) &= \frac{C(r)^2}{r^2},
\end{align}
and the initial profile of $\hat \Delta^r$ has been set according to Eq.~\eqref{eq:DeltaDefinition}. 

The lapse function $\alpha$ is initialized by imposing that $\tder K = 0$ at $t = 0$, therefore we integrated numerically the equation
\begin{multline}
	\rder^2 \alpha = \bigl( \rder \alpha \bigr) \biggl[\frac{\rder a}{2a} - \frac{\rder b}{b} - 6 \frac{\rder \psi}{\psi} - \frac{2}{r} \biggr] + \psi^4 \alpha a \frac{K^2}{3} + \\
	+ 4 \pi \alpha \psi^4 a (\mathcal E + S_a + 2S_b)
\end{multline}
together with Eq.~\eqref{eq:InitialConstraints psi}-\eqref{eq:InitialConstraints}. 
To solve the equations for the initial profile we used a shooting procedure starting the numerical integration from the origin and moving outwards. We imposed regularity at $r = 0$ and the asymptotic behaviors
\begin{align}
	E^r &= \frac{Q_\infty}{\tilde{r}^2} + \OO \Bigl( \frac{1}{\tilde{r}^3} \Bigr), \notag \\
	\psi &:= e^{\chi} = 1 + \frac{M_\text{\tiny ADM}}{2 \tilde{r}}+\OO \Bigl( \frac{1}{\tilde{r}^2} \Bigr), \notag \\
	K &= \OO \Bigl( \frac{1}{\tilde{r}^3} \Bigr), \notag \\
	\varphi &= \frac{Q_\infty}{\tilde{r}} + \OO \Bigl( \frac{1}{\tilde{r}^3} \Bigr),\notag \\
	\alpha &= 1 - \frac{M_\text{\tiny ADM}}{\tilde r} + \OO \Bigl( \frac{1}{\tilde r^2} \Bigr),
	\label{eq:AsymptoticInitialFlat}
\end{align}
where $M_\text{\tiny ADM}$ is the ADM mass, and $Q_\infty = Q(r_\infty)$ is the electric charge computed at the outer boundary.  
We performed the numerical integration using the Runge-Kutta method at the fourth order of accuracy, and the Newton's method as a root-finding algorithm in the shooting procedure. At the end of the initialization process we computed the ADM mass and electric charge at the outer boundary.

The numerical grid extends from the origin up to $\frac{r}{M} = 40$, with a grid step $\frac{\Delta r}{M} = 0.005$. The CFL factor was ${\rm CFL} = 0.01$, and we integrated the equations up to $\frac{T}{M} = 24$. 

\subsubsection{Results of the simulations}

We performed the numerical integration of the evolution equations for different values of $q M \in [0, 10]$, and studied the BH formation by computing the position of the  apparent horizon, $r=r_\text{\tiny AH}$. For the cases in which the collapse has happened we computed the horizon charge as
\begin{equation}
	Q_\text{\tiny AH} = \left.r^2 b \sqrt{a} E^r e^{6\chi}\right|_{r=r_\text{\tiny AH}},
	\label{eq:HorizonCharge}
\end{equation}
and the horizon mass using the Christodoulou-Ruffini mass formula~\cite{Christodoulou:1971pcn} in the case of vanishing spin:
\begin{equation}
	M_{H} = M_{\text{irr}} + \frac{Q_\text{\tiny AH}^2}{4 M_{\text{irr}}},
	\label{eq:M_AH}
\end{equation}
where $M_{\text{irr}} = \sqrt{\frac{A_H}{16 \pi}}$ is the irreducible mass and $A_H$ is the apparent horizon area.

The use of these formulas for the horizon mass and charge is based on the assumptions that the end state of the possible gravitational collapse is described by the RN metric and the final configuration is approximatly stationary near the origin at $t = T$. The first assumption is guaranteed by the uniqueness of the RN solution in Einstein-Maxwell theory, while the second assumption is satisfied for the value of $T$ that we chose. 

\begin{figure}
	\centering
	\includegraphics[width = \columnwidth]{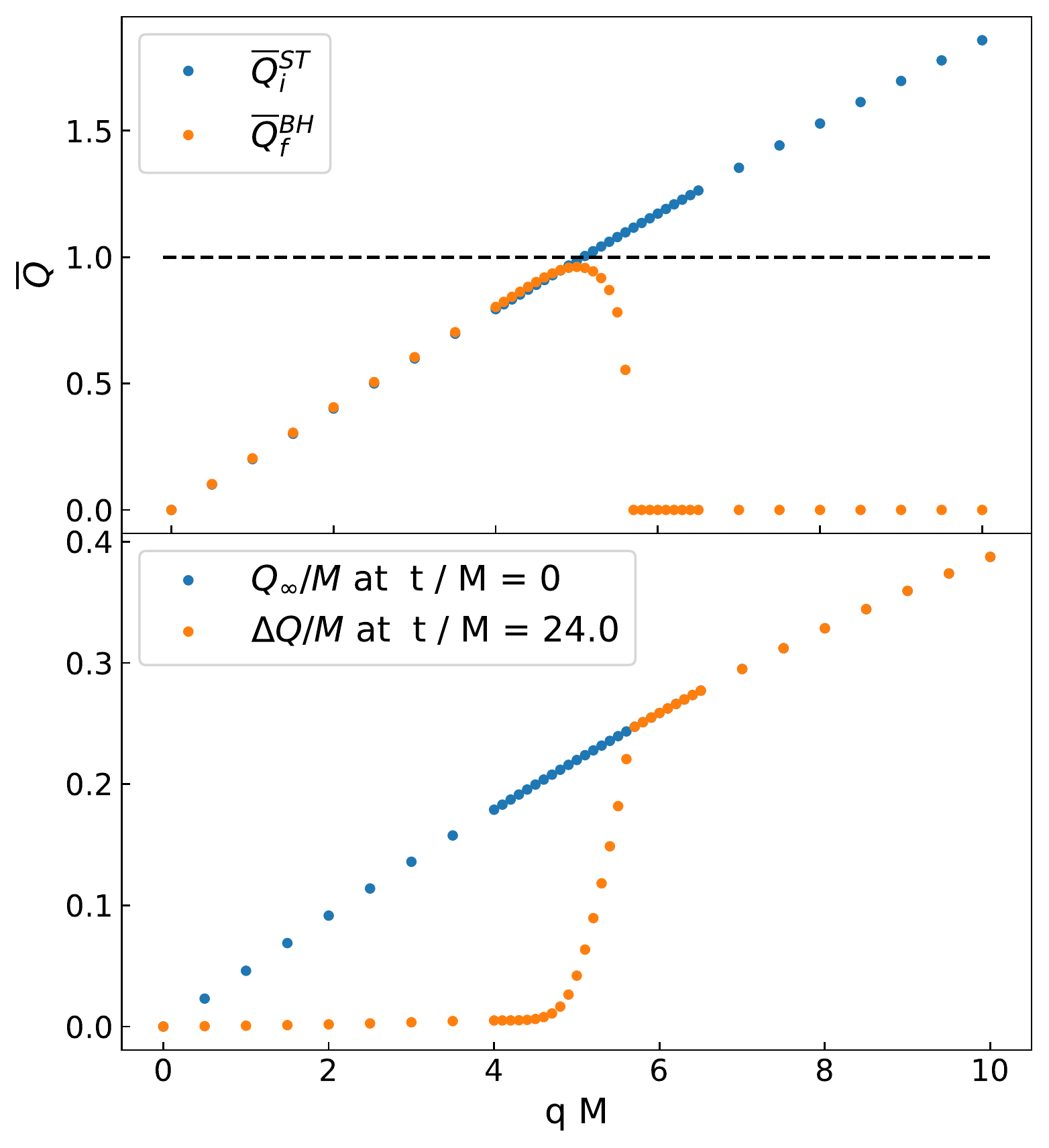}\\
	\caption{\textbf{Upper panel}: total charge-to-mass ratio of the spacetime at $t = 0$ (blue) and charge-to-mass ratio of the final BH at $t = T$ (orange) as functions of the charge $q$ of the initial ingoing wavepacket. \textbf{Lower panel}: total charge in the spacetime at $t = 0$ (blue) and amount of charge $\Delta Q$ outside the horizon at $t = T$ (orange) as functions of $q$. For low values of $q$ almost all the initial pulse collapses and forms the final BH, for $q M \sim 5$ the charge-to-mass ratio of the final BH reaches its maximum value and then decreases, due to the electromagnetic interaction that starts becoming dominant; finally, for $q M \gtrsim 5.65$ the gravitational collapse stops occurring, and there is no formation of a horizon. In the upper panel this condition conventionally corresponds to $\bar{Q}_f^\text{\tiny BH} = 0$ (i.e., $\Delta Q = Q_\infty$). 
	}
	\label{fig:flat_qbar}
\end{figure}

We then computed the initial (at $t=0$) charge-to-mass ratio of the full spacetime, $\bar Q_i^\text{\tiny ST} = \frac{Q_{\infty}}{M_\text{\tiny ADM}}$, and the charge-to-mass ratio of the final BH, $\bar Q_f^\text{\tiny BH} = \frac{Q_\text{\tiny AH}}{M_\text{\tiny AH}}$, at $t = T$. The results are shown in the upper panel of Fig.~\ref{fig:flat_qbar}. For $q M \lesssim 4.5$ almost all the scalar field present at the beginning of the simulation collapses and forms the final BH. When $q M \sim 5$ the final configuration is close to extremality but the BH remains subextremal. For $q M \gtrsim 5$ the charge-to-mass ratio of the spacetime at $t = 0$ exceeds unity and the electric forces start preventing the gravitational collapse: $\bar Q_{f}^\text{\tiny BH}$ rapidly decreases until $q M \sim 5.65$, where a BH stops forming. As a convention, in the plot we set $\bar{Q}_f^\text{\tiny BH} = 0$ for the cases in which a horizon does not form. 
The maximum value of the BH charge-to-mass ratio that we obtained in our simulations is $\bar Q_f^\text{\tiny BH} \sim 0.96$.

We also computed the amount of charge outside the final BH, $\Delta Q$, obtained by subtracting the horizon charge to the final charge computed at the outer boundary, and we compared it with the total electric charge at $t = 0$; the results are shown in the lower panel of Fig.~\ref{fig:flat_qbar}. For $q M \lesssim 4.5$ almost all the charge present in the initial pulse is enclosed in the horizon, then the amount of charge outside horizon starts increasing, and for $q M \gtrsim 5.65$ it coincides with the initial charge of the spacetime, since for these values of $q$ the electromagnetic interaction is strong enough to completely prevent the gravitational collapse.

\subsection{Collapse of the charged field towards a RN BH in Einstein-Maxwell theory}  \label{sec:collapseRN}

Next, we consider the collapse of the complex scalar field towards a RN BH within Einstein-Maxwell theory, attempting at overcharge it. As we shall show, not only does this allow to reach final BHs which are closer to extremality, but the process shows also superradiant amplification at full nonlinear level.

\subsubsection{Initial setup}

In this case the initial configuration of the system is given by a complex scalar field on a RN background. The parameters of the initial profile of $\xi$ are:
\begin{align}
 &B_0 = 0.002 \,,\quad k_0 M = 5\,, \notag \\
 &\sigma_\xi^2 M^2 = 2.5\,,\quad r_{0, \xi}/M = 20\,,
\end{align}
where in this case $M$ is set to be equal to the initial BH mass, $M_\text{\tiny BH} = M$, and all dimensionful quantities are measured in terms of $M$.

For this analysis we wish to construct a background configuration such that the mass and the charge of the central BH are fixed as $q$ varies. In order to achieve this we implemented a shooting algorithm that integrates Eqs.~\eqref{eq:InitialConstraints psi}-\eqref{eq:InitialConstraints} starting from the outer boundary and moving inward, and searches for the parameters $M_\text{\tiny ADM}$ (the ADM mass) and $Q$ in the asymptotic expansions 
\begin{align}
	E^r &= \frac{Q}{r^2} + \OO \Bigl( \frac{1}{r^3} \Bigr), \notag \\
	\psi &:= e^{\chi} = 1 + \frac{M_\text{\tiny ADM}}{2 r} - \frac{Q^2}{8 r^2} +\OO \Bigl( \frac{1}{r^3} \Bigr), \notag \\
	K &= \OO \Bigl( \frac{1}{r^3} \Bigr), \notag \\
	\varphi &= \frac{Q}{r} + \OO \Bigl( \frac{1}{r^3} \Bigr)\,,
	\label{eq:AsymptoticInitialRN}
\end{align}
such that the horizon charge and mass assume the required values. We used a precollapsed lapse~\cite{Alcubierre:2002kk} $\alpha = \frac{1}{\psi^2}$, and a conformal metric with $a = b = 1$, while the horizon mass was computed with the Christodoulou-Ruffini mass formula. After the initialization we extracted the total ADM mass and electric charge at the outer boundary. 

The BH initial charge-to-mass ratio was set to $\bar Q^\text{\tiny BH}_i = \frac{Q^\text{\tiny BH}_i}{M^\text{\tiny BH}_i} = \{ 0.9, 0.95, 0.99\}$. 
The numerical grid extends from the origin up to $\frac{r_\infty}{M} = 250$ with a grid step $\frac{\Delta r}{M} = 0.01$, and the CFL factor was ${\rm CFL} = 0.4$. The final time of integration was set to $\frac{T}{M} = 100$, which is sufficient to obtain an approximately stationary final configuration near the horizon. 

\subsubsection{Results of the simulations}

\begin{figure}
	\centering
	\includegraphics[width = \columnwidth]{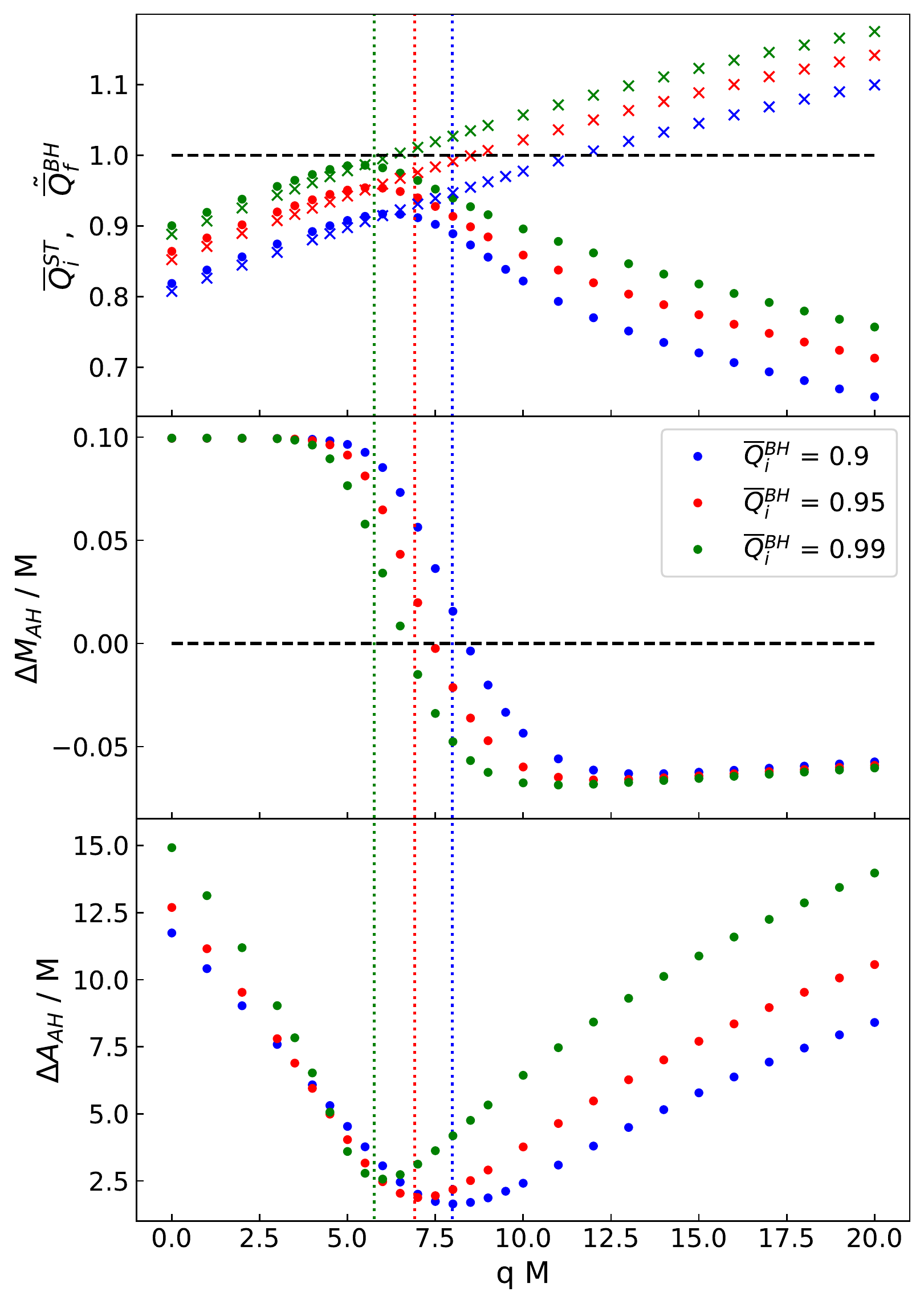}
	\caption{\emph{Cosmic censorship at play in Einstein-Maxwell theory.} \textbf{Upper Panel}: Charge-to-mass ratio $\bar{Q}^\text{\tiny BH}_f$ of the final BH (dots) and total charge-to-mass ratio $\bar{Q}^\text{\tiny ST}_i$ of the spacetime at the beginning of the simulations (crosses) for the collapse in Einstein-Maxwell theory. For low values of $q$ the incoming pulse is absorbed by the BH, and $\bar{Q}^\text{\tiny BH}_f$ increases with $q$, while for higher values of $q$ it decreases due to the electric repulsion and superradiance. Final configurations with an overcharged BH have never been produced. \textbf{Middle Panel}: Change of the BH mass during the simulation. For high values of $q$ superradiance takes place and extracts mass from the initial BH. \textbf{Lower Panel}: In all simulations the BH area increases, in agreement with the BH area law. The dotted lines in the three panels correspond to the threshold values for the superradiance condition summarized in Table~\ref{tab:SuperradiantThresholdsRN}.}
	\label{fig:RN_analysis}
\end{figure}

After the integration of the evolution equations for values of $q M \in [0, 20]$, we computed the charge-to-mass ratio of the final BH, $\bar{Q}^\text{\tiny BH}_f$. We plotted the results in the upper panel of Fig.~\ref{fig:RN_analysis}, where the dots represent $\bar{Q}^\text{\tiny BH}_f$ while the crosses represent the initial charge-to-mass ratio of the entire spacetime, $\bar{Q}^\text{\tiny ST}_i$. For low values of $q$ the charge carried by the complex field is smaller than its mass, the initial pulse is totally absorbed by the BH and $\bar{Q}^\text{\tiny BH}_f$ is smaller than $\bar{Q}^\text{\tiny BH}_i$. As $q$ increases the final charge-to-mass ratio increases, then reaches a maximum and starts decreasing, without producing overcharged final configurations. 
In this experiment, the maximum charge-to-mass ratio of the final BH achieved in our simulation is $\bar{Q}^\text{\tiny BH}_f\sim 0.986$.

This is not only due to the increasing electromagnetic repulsion that overcomes the gravitational attraction, but also to mass and charge extraction due to superradiance~\cite{Brito:2015oca}. In fact for sufficiently high values of the parameter $q$ the BH mass decreases during the evolution, as it can be seen from the middle panel of Fig.~\ref{fig:RN_analysis}, where we show the behavior of the difference between the final and the initial BH mass as a function of $q$. 

For a monochromatic test field on a RN background the superradiance condition is~\cite{Brito:2015oca}
\begin{equation}
	\omega < q \Phi_H,
	\label{eq:SuperradianceCondition}
\end{equation}
where $\omega$ is the wave frequency and $\Phi_H$ is the horizon electric potential. Therefore at a fixed frequency the condition \eqref{eq:SuperradianceCondition} is met for values of $q$ which are above the threshold $q_{\rm th} = \frac{\omega}{\Phi_H}$. Since we are not considering a monochromatic test field the superradiance condition is more involved, because the initial wavepacket contains both frequencies that satisfy Eq.~\eqref{eq:SuperradianceCondition} and higher frequencies which are instead absorbed by the BH. 
Nonetheless, we made an estimate of the threshold value $q_{\rm th}$ using $\omega = k_0$, where $k_0$ is the frequency in the initial profile of $\xi$, and the horizon electric potential of a RN BH $\Phi_H = \frac{Q_\text{\tiny AH}}{R_{H}}$, where $R_H$ is the horizon areal radius; the results are summarized in Table~\ref{tab:SuperradiantThresholdsRN}. As we can see the threshold values that we obtained are compatible with the behaviors in the middle panel of Fig.~\ref{fig:RN_analysis}, since they fall in the region where the difference between the final and initial BH mass is decreasing. Furthermore, the threshold value of $q$ decreases with $\bar Q^\text{\tiny BH}_i$, as expected. 
It is worth mentioning that the energy of the initial complex field is $\sim 0.1M$, so backreaction is relevant and the expectation from linear perturbation theory are only indicative.
Nonetheless, by comparing the top and middle panels in Fig.~\ref{fig:RN_analysis}, it is interesting to notice that the maximum of the final charge-to-mass ratio roughly corresponds to the BH mass extraction, suggesting that (nonlinear) superradiance plays an important role in preserving the cosmic censorship in Einstein-Maxwell theory. We will come back to this point later when performing a similar gedankenexperiment in Einstein-Maxwell-scalar theory.

\begin{table}
	\centering
	\begin{ruledtabular}
		\begin{tabular}{ddd}
			\multicolumn{1}{c}{$\bar{Q}_i^\text{\tiny BH}$}  &  \multicolumn{1}{c}{$\Phi_H$}  &  \multicolumn{1}{c}{$q_{\rm th} M$} \\
			\colrule
			0.9 & 0.63 & 8.0 \\
			0.95 & 0.72 & 6.9 \\
			0.99 & 0.87 & 5.8 \\
		\end{tabular}
	\end{ruledtabular}
	\caption{Estimates of the threshold values of the parameter $q_{\rm th}$ from the superradiance condition \ref{eq:SuperradianceCondition}. For $\Phi_H$ we used the horizon electric potential of a RN BH, $\Phi_H = \frac{Q_\text{\tiny AH}}{R_{H}}$, and for $\omega$ we used the frequency $k_0$ in the initial profile of the complex scalar field.}
	\label{tab:SuperradiantThresholdsRN}
\end{table}

Finally, in order to check the behavior of the entropy, in the lower panel of Fig.~\ref{fig:RN_analysis} we show the difference between the final and initial BH area. We can see that this value is always positive, in agreement with the BH area law in GR.

\subsection{Collapse of charged field towards a RN BH in nonminimally-coupled Einstein-Maxwell-scalar theory} \label{sec:scalarization}

Let us now move to our main analysis, which focuses on the collapse in Einstein-Maxwell-scalar theory with nonminimal couplings. 
We choose the simplest coupling that gives rise to spontaneous scalarization, $F[\phi]=1-\lambda\phi^{2}$ with $\lambda<0$. This provides a negative effective mass squared in the scalar perturbations, triggering a tachyonic instability of the RN BH. As a result of the instability, the BH scalarizes and a real scalar field profile forms around it.

\subsubsection{Static scalarized BHs}
Before performing numerical simulations,
we construct the scalarized charged BH solution assuming zero complex scalar field and a static spherically symmetric metric:
\begin{align}
ds^2 = - \Bigl( 1 - \frac{2m(R)}{R} \Bigr) e^{-2 \delta (R)} dt^2 + \frac{dR^2}{1 - \frac{2m(R)}{R}} + R^2 d\Omega^2\,, \label{static}
\end{align}
where $R$ is the areal radius, $m(R)$, and $\delta(R)$ are the metric functions.
Due to spherical symmetry, the only nonvanishing Maxwell equation can be directly integrated:
\begin{align}
\partial_{R}A_{t}(R)=\tilde{Q}\frac{e^{-\delta(R)}}{R^{2}(1-\lambda \phi(R)^{2})}\,,
\label{eq:StaticAtDer}
\end{align}
where $\tilde{Q}$ is the charge excluding the effect of the real scalar field [see Sec.~\ref{sec:ElectricCharge}].
By expanding around the BH horizon $R=R_{\rm H}$, 
we obtain 
\begin{align}
m(R)=\frac{R_{\rm H}}{2}+\frac{\tilde{Q}^{2}(R-R_{\rm H})}{2R_{\rm H}^{2}(1-\lambda\phi_{H}^{2})}
+\mathcal{O}(R-R_{\rm H})^{2}\,,
\end{align}
where $\phi_H=\phi(R_H)$ is the scalar field on the horizon.
Using a shooting method for finding $\phi_{H}$ with boundary condition $\phi(r\to \infty)=0$, 
we obtain the scalarized charged BH solution (see also Refs.~\cite{Herdeiro:2018wub, Fernandes:2019rez} where an equivalent computation has been performed). 
\begin{figure}
	\centering
	\includegraphics[width = \columnwidth]{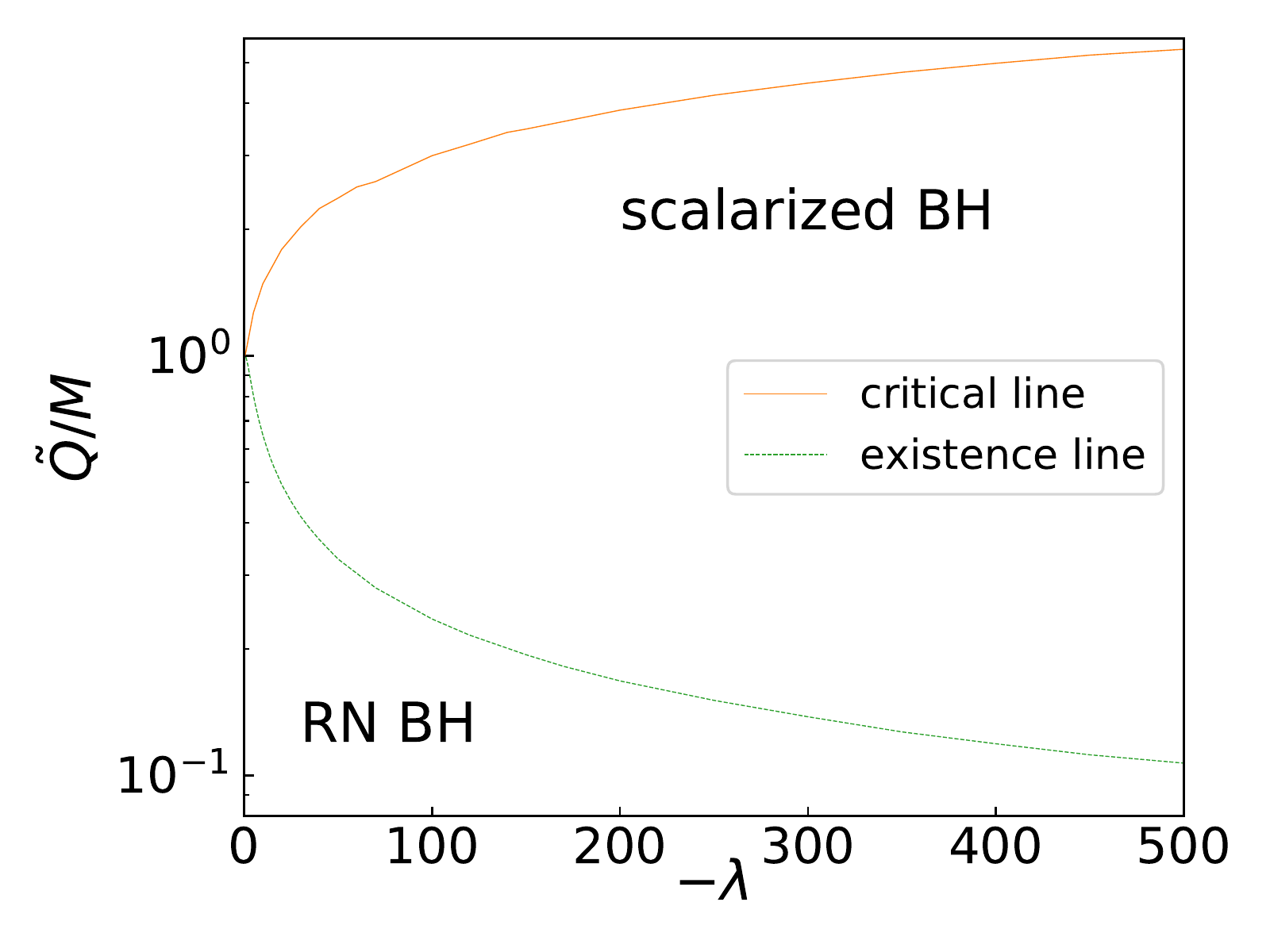}
	\caption{Domain of the existence of nodeless scalarized solutions in Einstein-Maxwell-scalar theory with $F[\phi]=1-\lambda\phi^{2}$. See also Ref.~\cite{Fernandes:2019rez} for an equivalent domain plot.}
	\label{fig:phase_diagram}
\end{figure}
In Fig.~\ref{fig:phase_diagram} we present the domain of the existence of nodeless scalarized solutions in this theory.
The existence line is the threshold for the stability of the RN BH, whereas the solutions on the critical line are singular at the horizon.
As shown in Fig.~\ref{fig:phase_diagram}, for a given value of $\lambda < 0$, scalarized BHs in this theory can exist in a certain range of charge-to-mass ratio and their maximum value of $\tilde{Q}/M$ can exceed the RN bound.

\subsubsection{Challenging the Cosmic Censorship~I: dynamical formation of scalarized charged BHs} \label{sec:Scalarization}

Let us move to study the dynamical formation of overcharged BHs in the presence of a nonminimal coupling. We set the coupling parameter to $\lambda = -500$ in such a way that the dynamics of the spontaneous scalarization is sufficiently fast and the computational cost of the simulation is moderate.

The setup of our gedankenexperiment is the following. We shall initially throw a small real scalar field onto a RN BH in a region of the parameter space in which the BH is unstable and scalarizes. We then throw a second wavepacket (this time made of a charged scalar field) which reaches the BH on longer time scales, i.e. when the BH is reaching a stationary configuration.
 Given the separation of scales, our setup is similar to trying to overcharge a hairy charged BH form the onset.

Thus, we wish to construct the initial configurations in such a way that the complex scalar field reaches the horizon sufficiently after the real scalar field. To this aim we use the same parameters as the previous analysis for the initial profile of $\xi$ and we initialized $\phi$ and $\Pi$ to
\begin{align}
	\phi(r, t = 0) &= A_0 \exp\left[{-\frac{(r-r_0)^2}{\sigma_0^2}}\right], \notag\\
	\Pi(r, t = 0) &= 0,
	\label{eq:InitialPhi}
\end{align}
where $A_0 = 0.0003$, $r_0/M = 10$ and $\sigma_0/M = \sqrt{8}$.  This initial profile coincides with the one used in Ref.~\cite{Herdeiro:2018wub}. Note that the amplitude of $\phi$ can be small since, owing to the tachyonic instability, the real scalar field initially grows exponentially during scalarization.
The initialization procedure is the same as in the previous section, with the difference that now the equations contain also the terms depending on $\phi$ as well as the corresponding dynamical equation for it. Initially, the real scalar field has neglibible support near the BH so we can consider the latter to be initially described by the RN metric. 
The grid parameters, the timestep, and the end time of the simulations are set to the same values as in the previous section.

During the evolution (and before the charged wavepackets reaches the horizon) we obtain a stable hairy BHs with nonvanishing profiles of the real scalar field. To compute the mass of the scalarized BHs we cannot use Eq.~\eqref{eq:M_AH}, since it is based on the hypothesis that the BH is described by the RN metric. An alternative strategy for extracting the mass could be to integrate the evolution equations for longer times, in such a way that the real scalar field profile of the final BH has reached a region of the spacetime large enough to compute the ADM mass explicitly. However this procedure is computationally expensive, since it requires large numerical grids and larger integration times.
We instead check that at $t = T$ the system has reached its final configuration near the horizon while the contribution from the complex scalar field can be neglected. In this case we can use the horizon data to construct a static scalarized BH solution from which we can then compute the ADM mass. This procedure heavily reduces the computational cost since it does not require to evolve the full system of equations for very long times.

The stationary configuration can be solved as previously explained (see Ref.~\cite{Herdeiro:2018wub} for details), using the ansatz~\eqref{static}.
In the integration of the equations the horizon areal radius $R_H$ and the horizon electric charge $\tilde Q_\text{\tiny AH}$ are taken from the numerical evolution at $t = T$, while $\phi(R_H)$ and $\delta(R_H)$ are found with a shooting procedure. We used the Newton's method as a root-finding algorithm. Since the scalarized solution is not unique, we initialized $\phi(R_H)$ using the end state of the evolution, in order to obtain the required profile of the real scalar field. 

We then computed the scalar charge $D$ as 
\begin{equation}
	D = -r^2 \frac{d \phi}{d R}\bigg\rvert_{R = R_\infty},
	\label{eq:ScalarCharge}
\end{equation}
where $R_\infty$ is the areal radius at the outer boundary, and the ADM mass as~\cite{Herdeiro:2018wub}
\begin{equation}
	M = m(R_\infty) + \frac{\tilde Q_\text{\tiny AH}^2 + D^2}{2 R_\text{\tiny AH}}\,.
	\label{eq:MStatic}
\end{equation}

\begin{figure}
	\centering
	\includegraphics[width = \columnwidth]{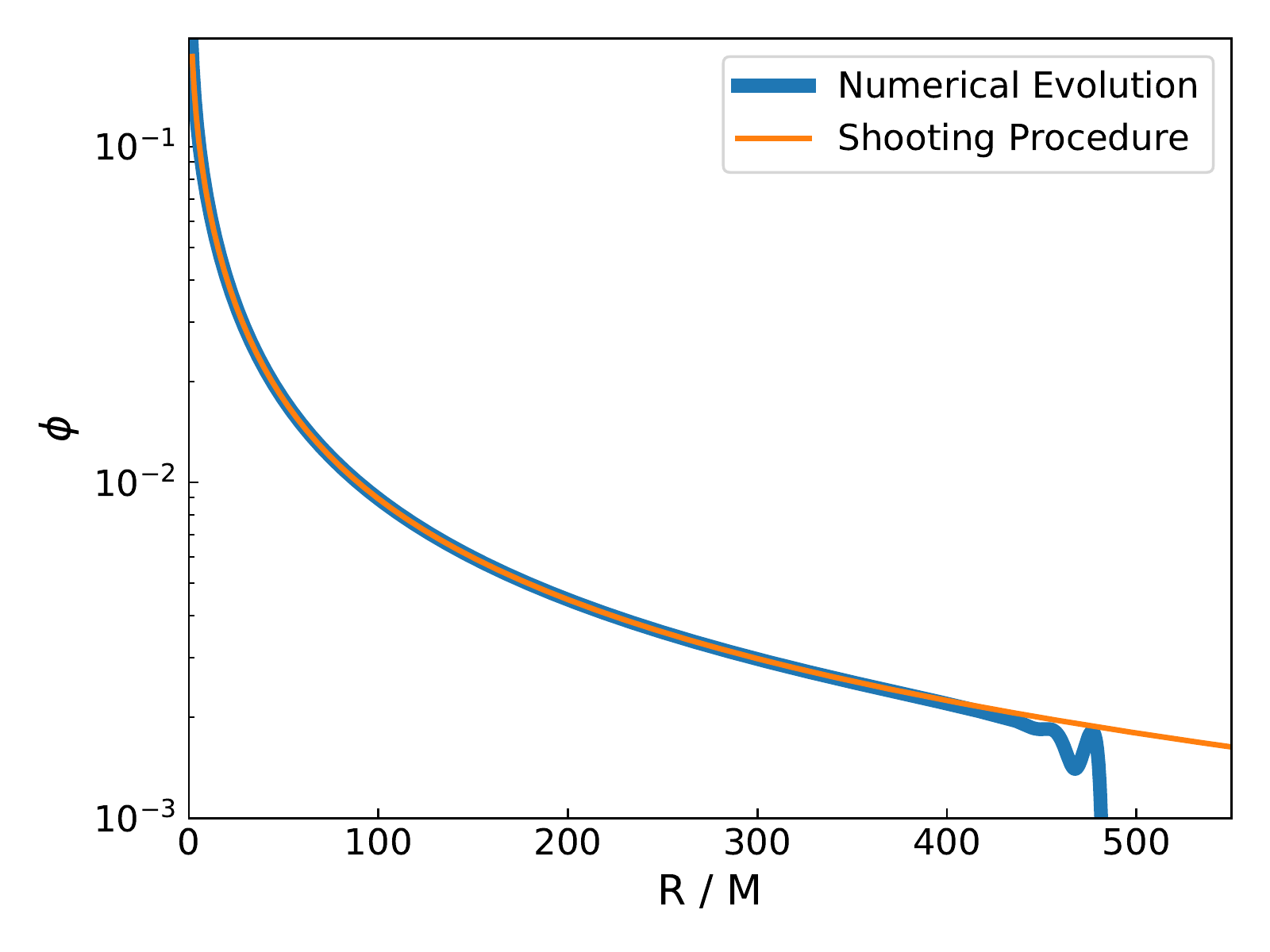}
	\caption{Profiles of the real scalar field obtained from the numerical integration at $T = 500M$ (blue) and from the shooting procedure extracting the parameters at $t = 100M$ (orange). The static scalarized solution is an excellent approximation of the end state of the evolution for $R<T$, as expected.}
	\label{fig:ScalarFieldComparison}
\end{figure}

In order to show the accuracy of this procedure, we performed a numerical integration of the field equations in the case of $\bar{Q}_i^\text{\tiny BH} = 0.9$ until $t = T = 500M$, using a grid that extends up to ${r_\infty} = 550{M}$; we then compared the profile of the scalar field at the final time with the static scalarized solution computed extracting the parameters at $t = 100M$. The results are shown in Fig.~\ref{fig:ScalarFieldComparison}, where we can see that the static solution accurately reproduces the end state of the numerical evolution, and the integration time $T = 100M$ is sufficient to obtain reliable estimates of the mass and electric charge of the final scalarized BH. 

Once the mass has been extracted we can compute the charge-to-mass ratio of the final BH using the definition of the charge, Eq.~\eqref{eq:TildeQdef}.

\begin{figure}
	\centering
	\includegraphics[width = \columnwidth]{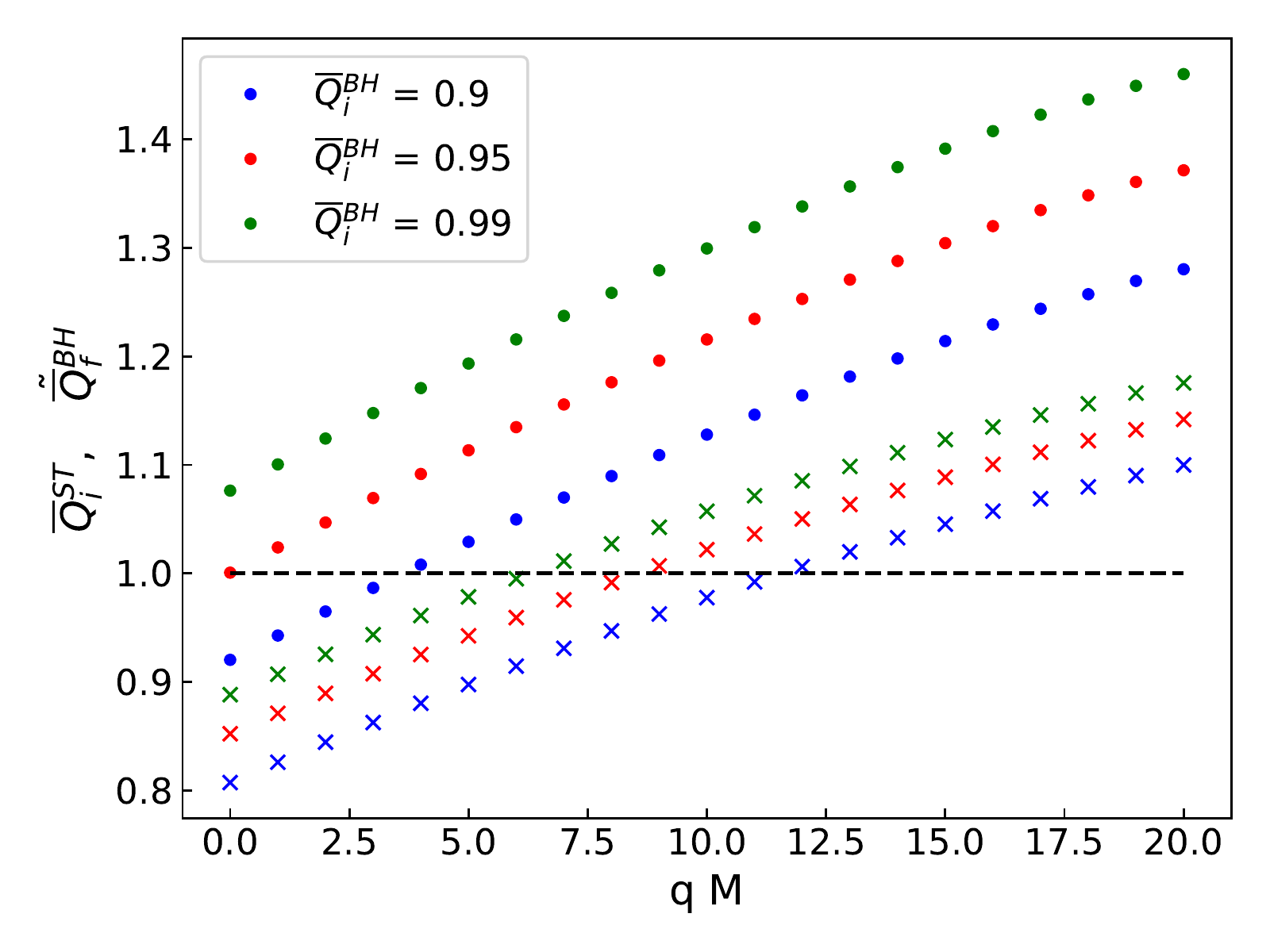}
	\caption{Charge-to-mass ratio $\tilde{\bar{Q}}^\text{\tiny BH}_f = \frac{\tilde{Q}_f^\text{\tiny BH}}{M_\text{\tiny BH}}$ of the final BH (dots) and total charge-to-mass ratio $\bar{Q}^\text{\tiny ST}_i$ of the spacetime at the beginning of the simulations (crosses) for the collapse in Einstein-Maxwell-scalar theory. This plot should be compared with the top panel of Fig.~\ref{fig:RN_analysis}. In this case overcharged configurations are formed; this is due to the presence of the nonminimal coupling that quenches the electromagnetic interaction and allows to enclose a large amount of charge within the horizon.}
	\label{fig:barQscalarization}
\end{figure}

One of our main results is shown in Fig.~\ref{fig:barQscalarization}, where one can see that overcharged configurations are generically produced. 
Nonetheless, the endstate of the collapse is always a (scalarized) BH and no naked singularities were produced in our gedankenexperiments. This suggests that the cosmic censorship is not a prerogative of GR but is also at play in Einstein-Maxwell-scalar theory. We will further discuss this point in Sec.~\ref{sec:CCCII}.

\begin{figure}
	\centering
	\includegraphics[width = \columnwidth]{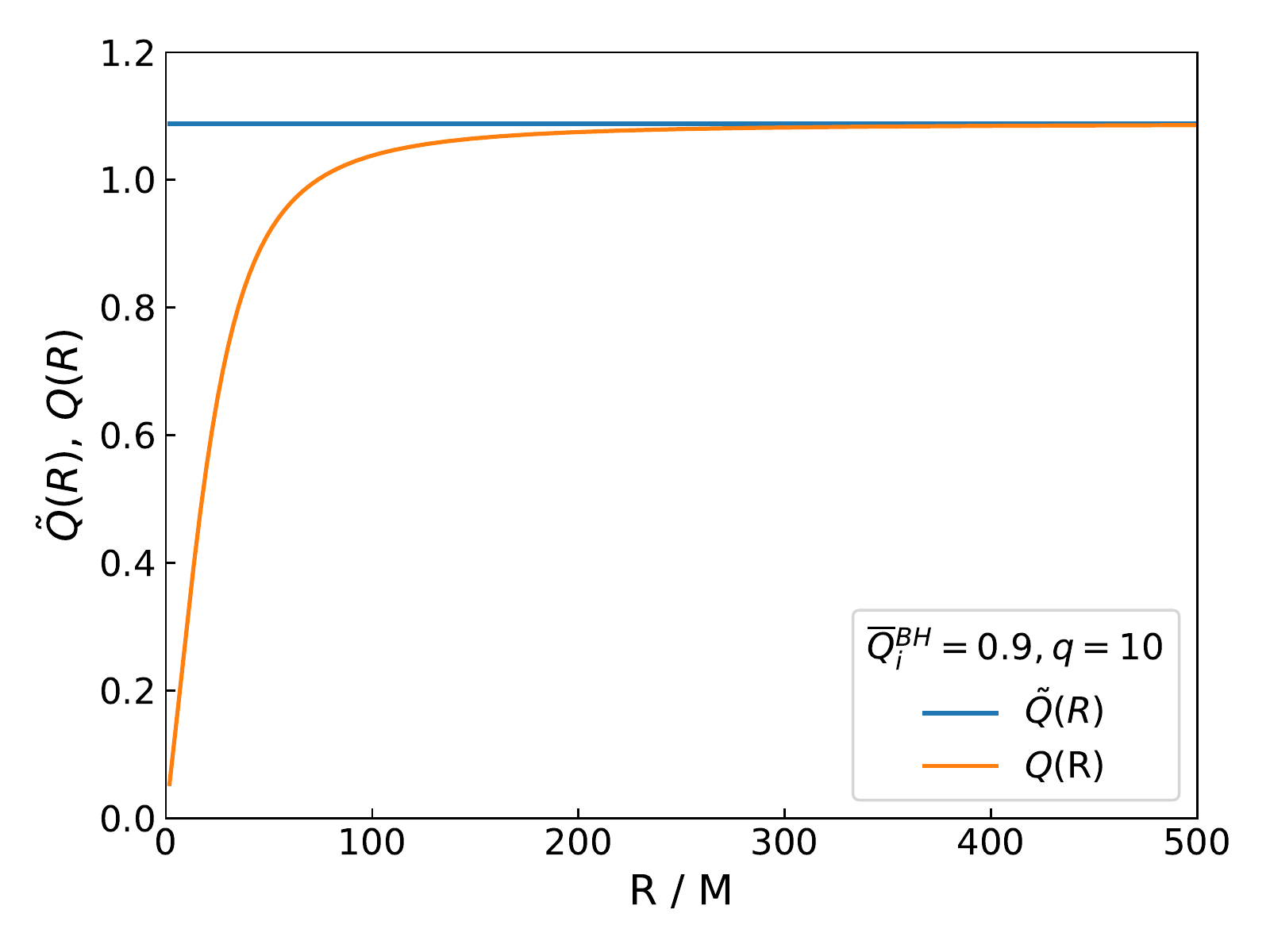}
	\caption{Charge enclosed in the 2-sphere of areal radius $R$ for a static scalarized solution computed at the end of the numerical evolution. $\tilde Q$ accounts only for the contribution of the charged field, and is constant in $R$ when the complex scalar $\xi$ is absent. On the other hand $Q$ can be seen as a parameter that measures the ``strength'' of the electromagnetic interaction, and it decreases near the horizon for a scalarized configuration.}
	\label{fig:ChareComparison}
\end{figure}

For the static solution that we constructed the profile of the electric field is given by Eq.~\eqref{eq:ErEMS}; in this expression the charge $\tilde Q$ accounts only for the contribution from the charged fields (see discussion in Sec.~\ref{sec:ElectricCharge}), and $\phi$ appears at the denominator via the coupling function, which is positive. Therefore the appearance of overcharged solutions may be explained by the action of the real scalar field that quenches the electric interaction, allowing to construct configurations in which a large amount of charged matter is confined within the horizon due to gravitational attraction. In this sense the electric charge $Q = \frac{\tilde Q}{F[\phi]}$ can be interpreted as a parameter that represents the ``strength'' of the electromagnetic interaction. In Fig.~\ref{fig:ChareComparison} we show the charge enclosed in the 2-sphere of areal radius $R$ for a static scalarized configuration, using the two definitions \eqref{eq:Qdef} and \eqref{eq:TildeQdef}; as we can see $\tilde Q$ is constant, while $Q$ decreases near the horizon due to the presence of the real scalar field.

Finally, it is worth mentioning that for high values of $\bar{Q}_i^\text{\tiny BH}$ overcharged final configurations are produced even when $q = 0$ and the field $\xi$ does not carry any contribution to the BH charge. This happens because part the mass of the BH is ejected in a scalar spherical wave during the scalarization process

\subsubsection{Induced descalarization of hairy BHs by absorption of opposite-charged wavepackets}

Next, we study the possibility of forming a RN BH from a previously scalarized configuration.

As we can see from Fig.~\ref{fig:phase_diagram} for low values of the BH charge-to-mass ratio the system does not admit scalarized configurations. Our objective is to dynamically produce a RN BH from a previously scalarized one. To do this we will start from a RN BH and induce the spontaneous scalarization with a perturbation of the real scalar field; once the central BH has reached a stable configuration, we will send a pulse of the complex scalar field with opposite charge in such a way that the final BH has charge close to zero and it is forced to descalarize. 

\begin{figure}
	\centering
	\includegraphics[width = \columnwidth]{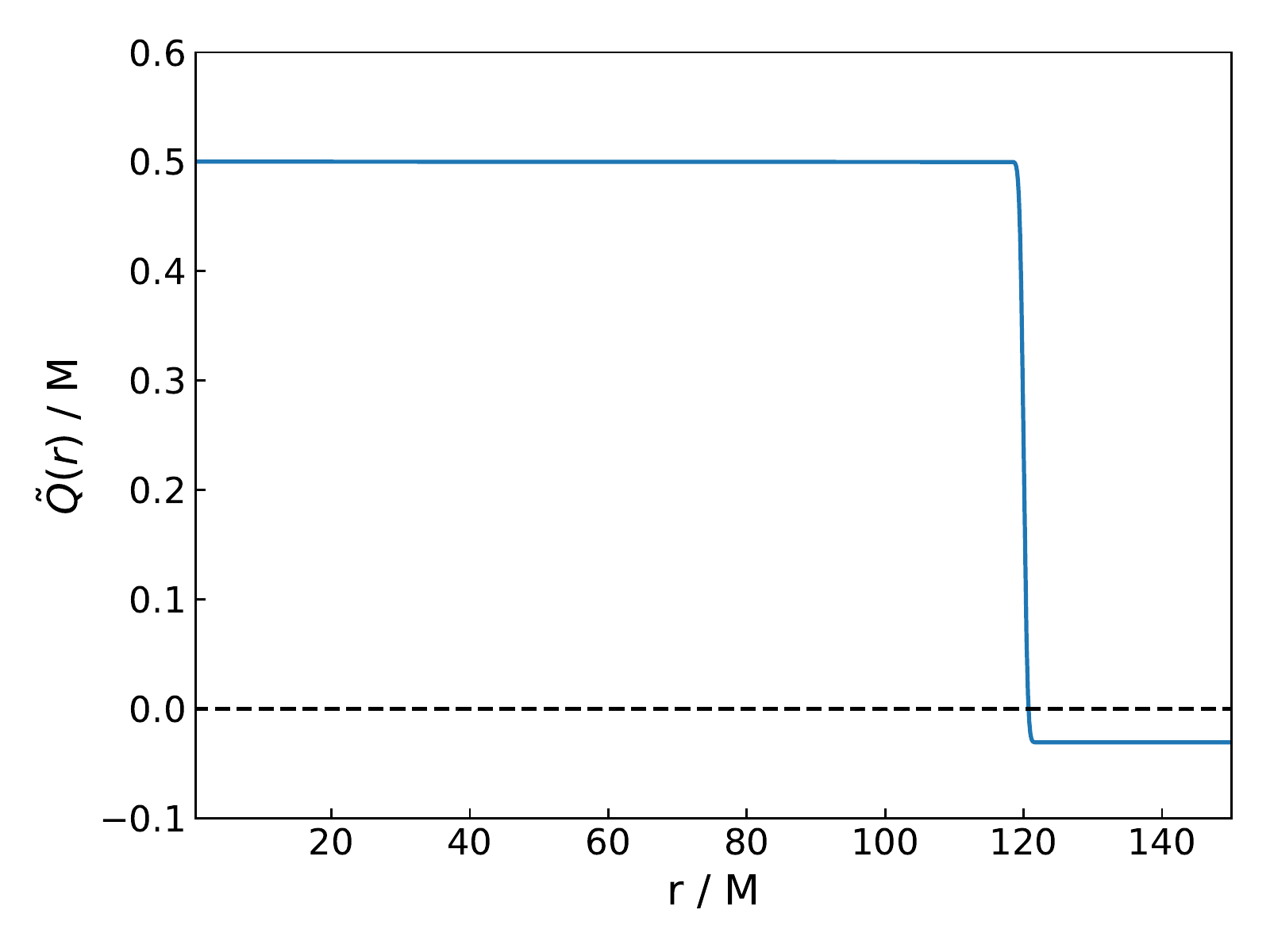}
	\caption{Profile of electric charge contained in the 2-spheres of radius $r$ at $t = 0$. The charge carried by the complex scalar field is such that the total charge in the spacetime is close to zero. In this way when the pulse is absorbed by the BH, the system will be in a region of the parameter space in which no scalarized solutions exist.
	}
	\label{fig:DescalarizationChargeProfile}
\end{figure}

We construct the initial configuration using the same shooting procedure described before, setting the BH mass to $M_i^\text{\tiny BH} = M$ and the initial charge-to-mass ratio to $\bar{Q}_i^\text{\tiny BH} = 0.5$. For the real scalar field we consider the profile in Eq.~\eqref{eq:InitialPhi}, while for the complex scalar field we exchange the real and the imaginary parts in Eq.~\eqref{eq:InitialXi} (in order to have a wavepacket with opposite charge) and we set the parameters to
\begin{align}
 & B_0 = 0.0004 \,,\quad k_0 M = 5\,, \notag \\
 & \sigma_\xi^2 M^2 = 2.5\,,\quad r_{0, \xi}/M = 120\,.
\end{align}
We also set $q M = 20$. In this way we obtain an inward-moving, negatively charged initial profile for $\xi$, such that the total charge of the spacetime is close to zero. The profile of the electric charge contained in the 2-spheres of radius $r$ at $t = 0$ is shown in Fig.~\ref{fig:DescalarizationChargeProfile}.

For the numerical evolution we chose a grid that extends up to $r_\infty = 400M$, with a grid step $\Delta r = 0.01M$. The CFL factor was ${\rm CFL} = 0.5$ and the final integration time was $T = 240M$. 

\begin{figure}
	\centering
	\includegraphics[width = \columnwidth]{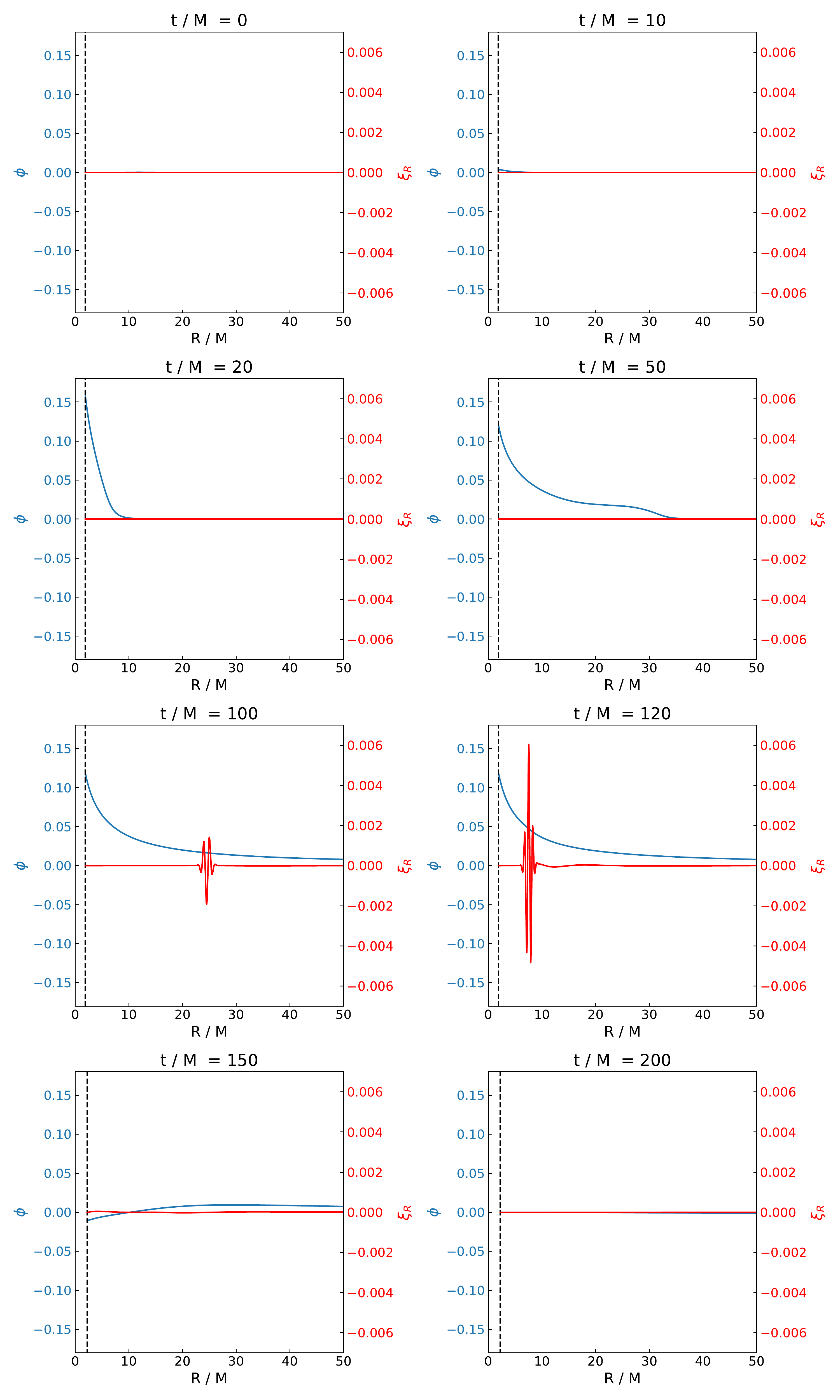}
	\caption{Snapshots of the evolution of the real scalar field $\phi$ (blue) and the real part of the complex scalar field $\xi$ (red) for the process of scalarization and subsequent descalarization of a RN BH. The black dashed line shows the position of the apparent horizon. Initially the complex scalar field does not affect the dynamics of the system and the perturbation of the real scalar field triggers the spontaneous scalarization of the BH, that reaches a stable configuration. Then, when the complex scalar field reaches the horizon it is absorbed by the BH, which descalarizes leaving a final RN BH.
	}
	\label{fig:DescalarizationSnapshots}
\end{figure}

In Fig.~\ref{fig:DescalarizationSnapshots} we show some snapshots\footnote{Some animations of this gedankenexperiment are available online~\cite{webpage}.} of the evolution of $\phi$ (in blue) and the real part of $\xi$ (in red).
As we can see in the first part of the evolution the BH is not affected by the complex scalar field and scalarizes reaching a stable configuration near in the central region. Later, the charged pulse reaches the horizon and is absorbed by the BH that, being in a region of the parameter space in which there is no stable scalarized solution, descalarizes leaving a final RN BH. 

\begin{figure}
	\centering
	\includegraphics[width = \columnwidth]{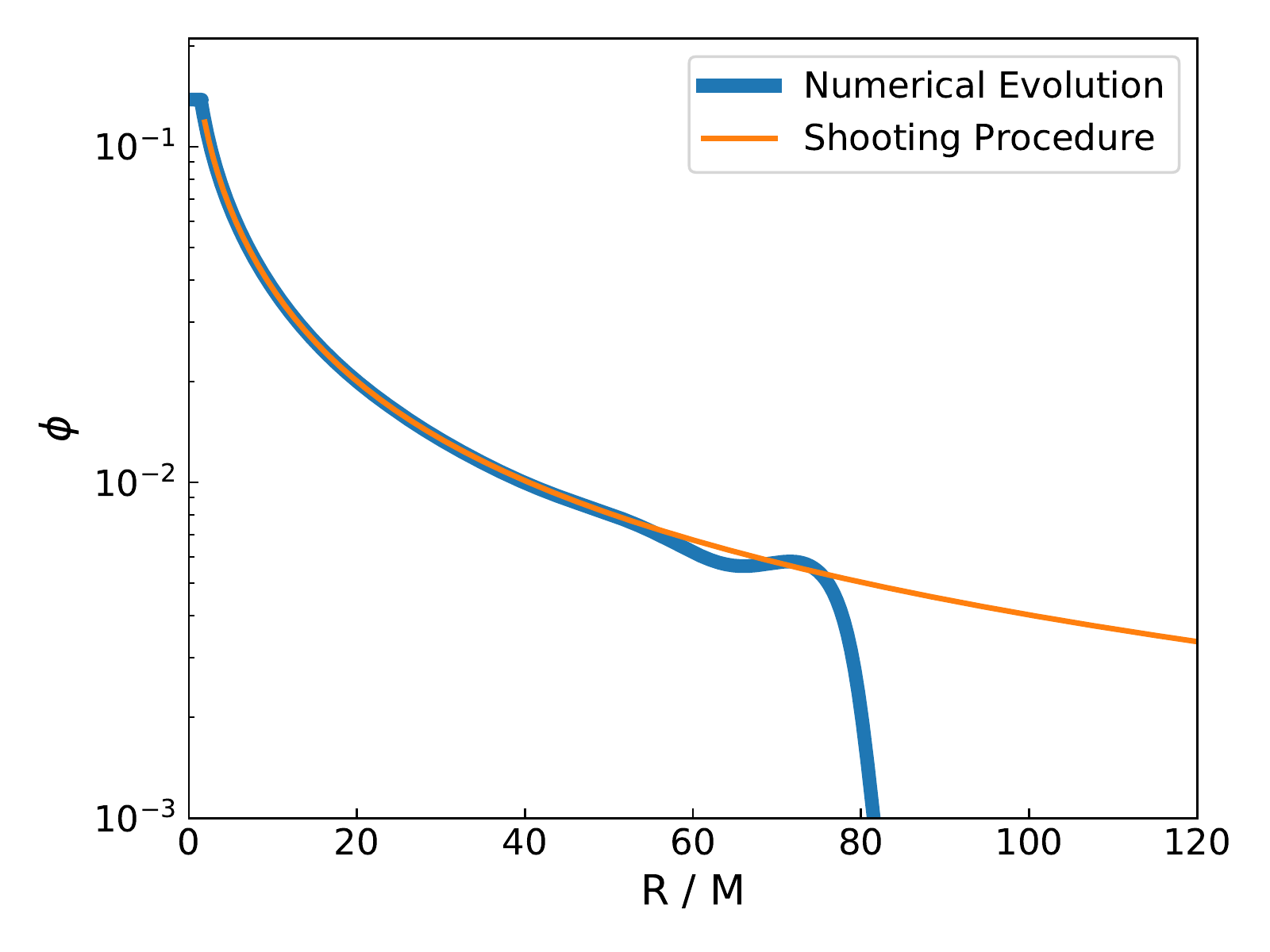}
	\caption{Comparison between the profile of the real scalar field at $T = 100M$ for the induced descalarization process (blue) and the static scalarized configuration obtained with the shooting procedure described in Sec.~\ref{sec:Scalarization} (orange). As we can see there is a good agreement between the two profiles, and we can consider that in the central region the scalarization process is completed.}
	\label{fig:DescalarizationStatic}
\end{figure}

To check that the BH at $t = 100M$ can be described by a scalarized solution, we compared the profile of the scalar field with the static scalarized configuration obtained using the shooting procedure described in the previous section. The result is shown in Fig.~\ref{fig:DescalarizationStatic}, where we can see that there is a good agreement between the two profiles. Thus we can assume that in the central region the scalarization process is completed, and that the subsequent part of the evolution shown in Fig.~\ref{fig:DescalarizationSnapshots} (i.e., $t\gtrsim 100M$) can be considered a descalarization process.

\subsubsection{Challenging the Cosmic Censorship~II: superradiantly-induced descalarization} \label{sec:CCCII}
From the results shown in Fig.~\ref{fig:barQscalarization} we observe that scalarized BHs beyond the RN bound can form dynamically and their final charge-to-mass ratio grows with the charge of the initial wavepacket $q$. On the other hand, the domain plot in Fig.~\ref{fig:phase_diagram} shows that, for a fixed value of $\lambda<0$, scalarized BHs can exist only below a critical value of the charge-to-mass ratio. Although the critical value is above unity and depends on $\lambda$, the situation is akin to the RN case. It is therefore natural to ask whether one can overcharge a \emph{scalarized} BH past its own extremality, possibly producing a naked singularity.
In this section we study this problem, showing that also in this case the superradiant extraction of the BH charge and mass plays a crucial role to bound the final charge-to-mass ratio below extremality.

To this purpose, we simulate the following process: we start with a RN BH and a small perturbation of the real scalar field so that the BH scalarizes; once the scalarization process has completed in a region sufficiently large around the horizon, a pulse of the complex scalar field interacts with the BH, and sets it to a new equilibrium state that we want to study. 

In this case the horizon electric potential that appears in Eq.~\eqref{eq:SuperradianceCondition} should be computed by integrating Eq.~\eqref{eq:StaticAtDer}, in which the coupling function appears at the denominator. Therefore in order to encounter the superradiant behavior for low values of $q$, we chose a small (negative) value of the coupling functions: $\lambda = -10$. This makes the initial scalarization time scale longer than in the $\lambda=-500$ case previously explored, so we need to throw the complex scalar field sufficiently later in order to make sure it interacts with the BH after the scalarization has completed. We therefore place the initial pulse of $\xi$ far from the origin, setting the initial profile of $\xi$ according to Eq.~\eqref{eq:InitialXi} with parameters
\begin{align}
 & B_0 = 0.0003 \,,\quad k_0 M = 2\,, \notag \\
 & \sigma_\xi^2 M^2 = 0.5\,,\quad r_{0, \xi}/M = 150\,.
\end{align}
This guarantees that, when the pulse of the complex scalar field reaches the BH, the scalarization process is completed in the horizon region. We also chose a smaller $k_0$ than in the previous case of standard Einstein-Maxwell theory in order for superradiance to occur at smaller values of $q$.

\begin{figure}
	\centering
	\includegraphics[width = \columnwidth]{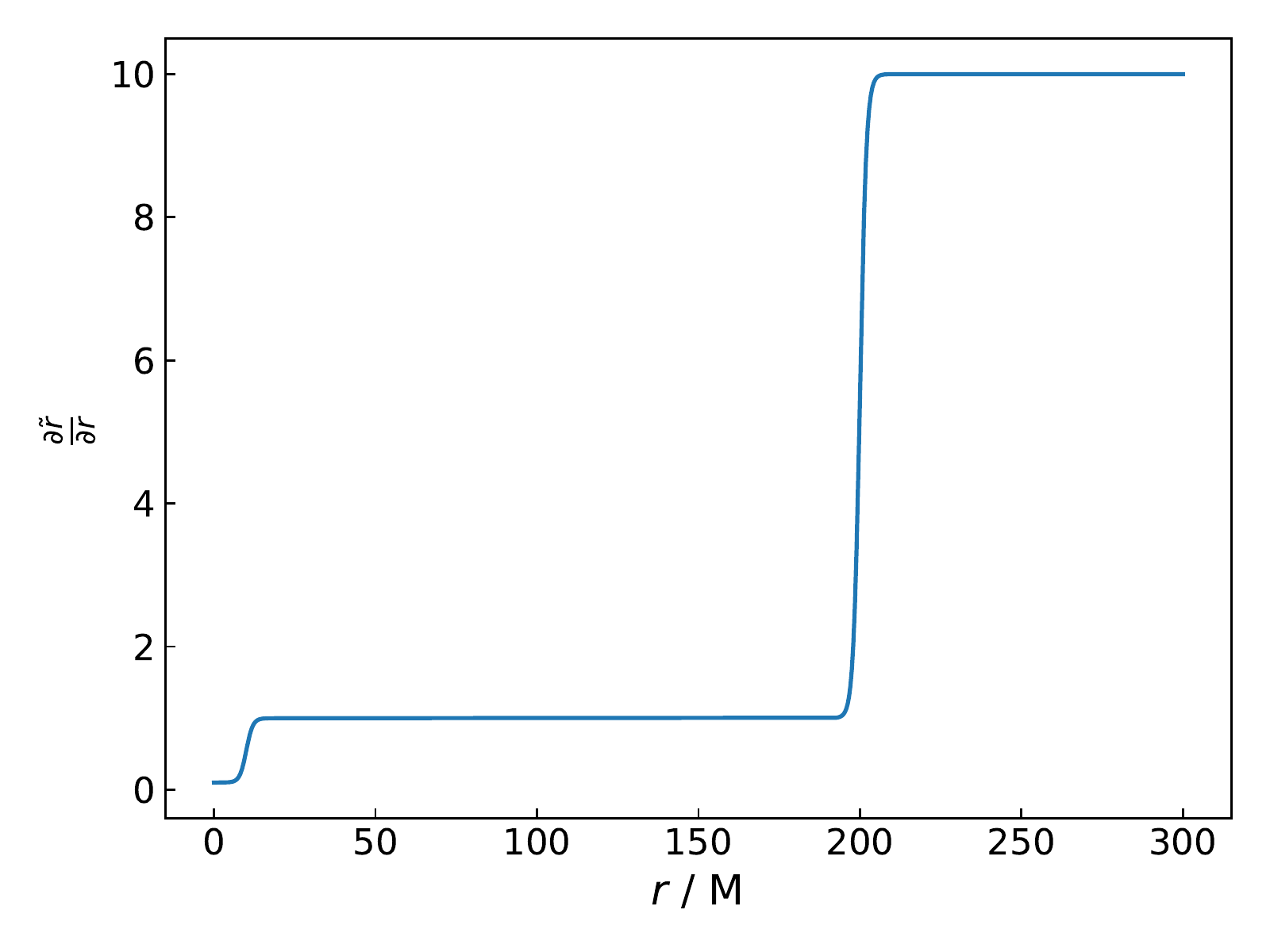}
	\caption{Derivative of the transformation $\tilde r = C(r)$ for the implementation of the nonuniform grid step. In the horizon region this transformation is analogous to the one used for the collapse on flat background, allowing larger grid steps without losing accuracy. Instead far from the origin small region in the coordinate $\tilde r$ are mapped into large region in $r$, allowing the use of a smaller numerical grid.}
	\label{fig:ScalarizedTilderDer}
\end{figure}

We implemented a nonuniform grid step in order to reduce the computational cost of the simulations. In particular the radial coordinate was transformed according to:
\begin{equation} 
	\begin{cases}
		\tilde r &= C(r) =  \eta_2 r + \frac{1 - \eta_1}{\Delta} \ln \Bigl( \frac{1 + e^{-\Delta (r - R_1)}}{1 + e^{\Delta R_1}} \Bigr) + \\
			 &+ \frac{1 - \eta_2}{\Delta} \ln \Bigl( \frac{1 + e^{-\Delta (r - R_2)}}{1 + e^{\Delta R_2}} \Bigr) \\
		\frac{\partial \tilde r}{\partial r} &= C'(r) = \eta_1 + \frac{1 - \eta_1}{1 + e^{-\Delta (r - R_1)}} + \frac{1 - \eta_2}{1 + e^{-\Delta (r - R_2)}}
	\end{cases}
	\label{eq:ScalarizedTilder}
\end{equation}
where again is understood that $r$ is the new coordinate and $\tilde{r}$ is the old one. We choose $\Delta = 1/M$, $\eta_1 = 0.1$, $\eta_2 = 10$, $R_1=10M$, and $R_2=200M$.
The profile of the derivative $C'(r)$ is shown in Fig.~\ref{fig:ScalarizedTilderDer}; for low values of $r$ this transformation is analogous to the one used for the collapse on flat background, while far from the origin large intervals in the coordinate $\tilde r$ are mapped into small intervals in $r$. In this way we can use a relatively large grid step without losing accuracy at the horizon, and we satisfy the condition that the signals do not reach the outer boundary even with a smaller numerical grid. 

To construct the initial configuration we used the same shooting procedure as for the other simulations described in this section, imposing the following asymptotic behaviors:
\begin{align}
	E^r &= \frac{Q_\infty}{\tilde{r}^2 \frac{\partial \tilde{r}}{\partial r}} + \OO \Bigl( \frac{1}{\tilde{r}^3} \Bigr), \notag \\
	\psi &:= e^{\chi} = 1 + \frac{M_\text{\tiny ADM}}{2 \tilde{r}}+\OO \Bigl( \frac{1}{\tilde{r}^2} \Bigr), \notag \\
	K &= \OO \Bigl( \frac{1}{\tilde{r}^3} \Bigr), \notag \\
	\varphi &= \frac{Q_\infty}{\tilde{r}} + \OO \Bigl( \frac{1}{\tilde{r}^3} \Bigr),\notag \\
	\label{eq:AsymptoticInitialScalarized}
\end{align}
where again $M_\text{\tiny ADM}$ is the ADM mass and $Q_\infty = Q(r_\infty)$.

The outer boundary was placed at $\frac{r_\infty}{M} = 250$, and the grid step was $\frac{\Delta r}{M} = 0.025$. The final time of integration was $\frac{T}{M} = 300$, and ${\rm CFL} = 0.05$. 

\begin{figure}
	\centering
	\includegraphics[width = \columnwidth]{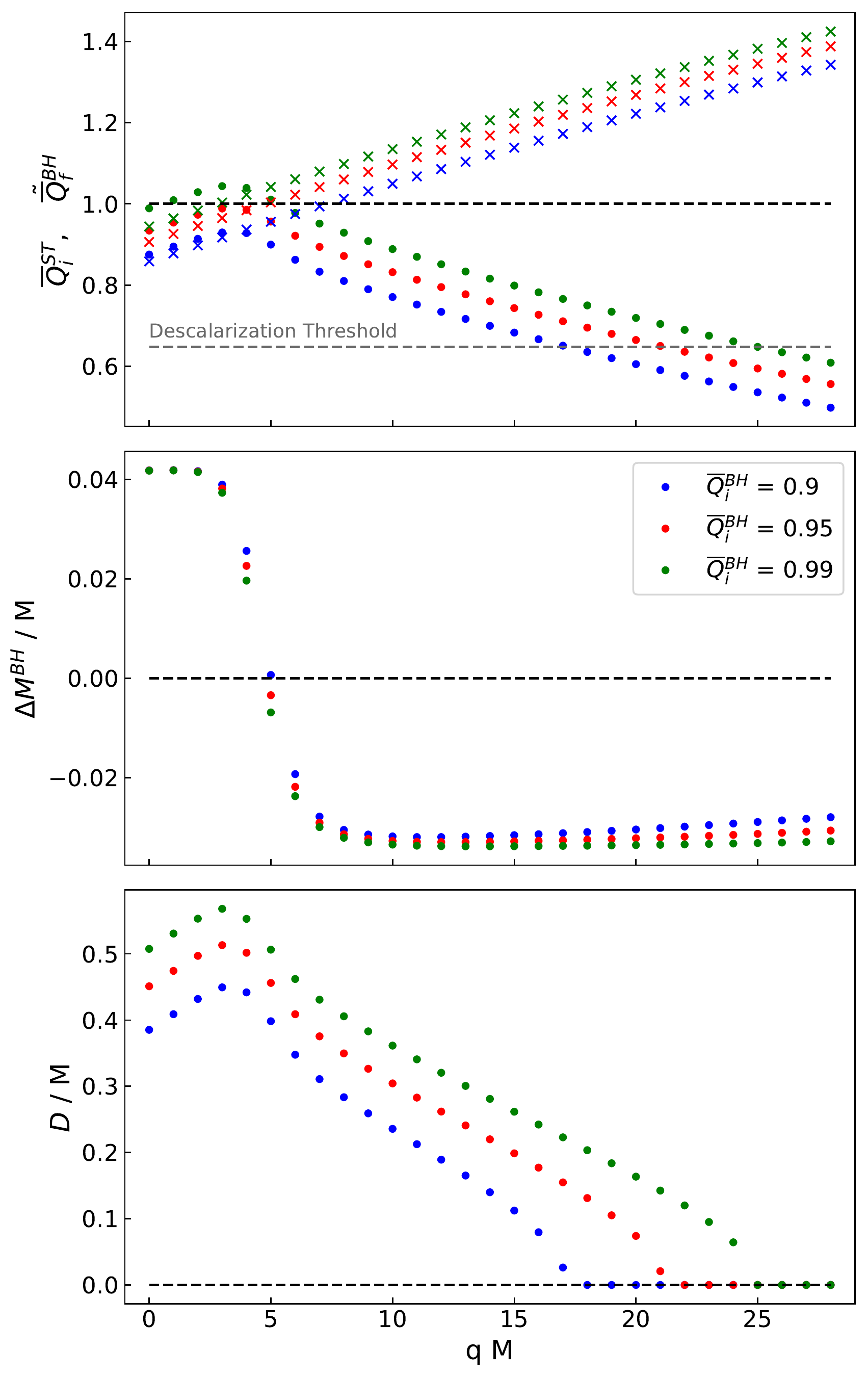}
	\caption{Results for the collapse of a complex scalar field in Einstein-Maxwell-scalar theory, with quadratic coupling and $\lambda = -10$. \textbf{Upper panel}: Charge-to-mass ratio $\bar{Q}^\text{\tiny BH}_f$ of the final BH (dots) and total charge-to-mass ratio $\bar{Q}^\text{\tiny ST}_i$ of the spacetime at the beginning of the simulations (crosses). \textbf{Middle panel}: mass difference between the final and the intermediate scalarized BH. \textbf{Lower panel}: scalar charge of the final BH. The charge-to-mass ratio of the final BH increases for low values of $q$, then it reaches a peak and starts decreasing. The negative $\Delta M^\text{\tiny BH}$ for high values of $q$ indicates the presence of superradiance. This mechanism can be efficient enough that the final charge-to-mass ratio falls below the threshold value for scalarization (grey dashed line in the upper panel), leading to the descalarization of the BH.}
	\label{fig:EMS_analysis_merged_D}
\end{figure}

We computed the final BH mass using the static scalarized solution that approximates the configuration of the system in the central region, and we studied the behavior of the charge-to-mass ratio of the final BH; the results are shown in the upper panel of Fig.~\ref{fig:EMS_analysis_merged_D}. As we can see the charge-to-mass ratio increases for small values of $q$, then reaches a peak and starts decreasing, as in the Einstein-Maxwell case. 
In the middle panel we show the mass difference between the final BH and the intermediate scalarized one. Interestigly, the mass of the final BH is smaller, showing that superradiance is at play also for the scalarized BH\footnote{Note that a linear study of superradiant scattering off a scalarized BH in Einstein-Maxwell-scalar theory is much more involved than in the RN case in Einstein-Maxwell theory, since electromagnetic and scalar perturbations are coupled to each other. Hence, in this case we do not have a prediction for the threshold value of $q$.}. Indeed, also in this case the maximum of the charge-to-mass ratio roughly corresponds to the onset of superradiance at nonlinear level. 
As for the collapse of the complex scalar field on a RN BH in Einstein-Maxwell theory, the extraction of charge is more efficient than the extraction of mass, so that $\tilde{\bar{Q}}_f^\text{\tiny BH}$ decreases. In other words, although the final charge-to-mass ratio can exceed the RN bound, it cannot grow indefinitely due to superradiance and reaches a maximum which is below the extremal value.

Interestingly enough, superradiance can be so efficient that the charge-to-mass ratio of the final BH can eventually cross the scalarization threshold (grey dashed line in the upper panel of Fig.~\ref{fig:EMS_analysis_merged_D}), leading to \emph{superradiantly-induced descalarization}. This can be clearly seen from the behavior of the final scalar charge $D$ (lower panel of Fig.~\ref{fig:EMS_analysis_merged_D}): for large values of $q$ the scalar charge goes to zero, indicating that the final BH has lost all its scalar hair. See~\cite{webpage} for some animations of these simulations.

\begin{figure}
	\centering
	\includegraphics[width = \columnwidth]{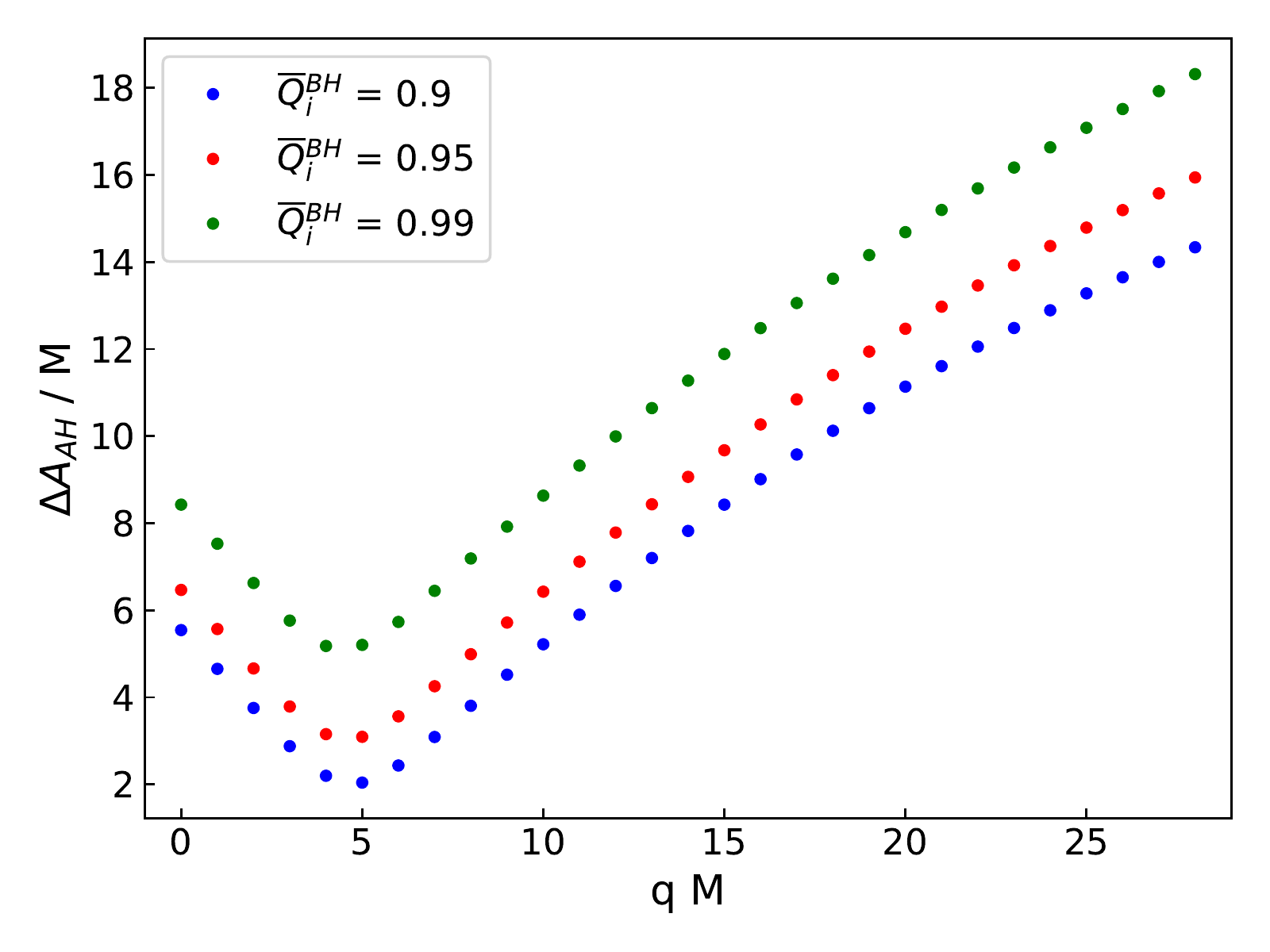}
	\caption{Difference between the final and initial BH area. As we can see the BH area always increases, in agreement with the area law.}
	\label{fig:EMSDeltaA}
\end{figure}

Finally, in Fig.~\ref{fig:EMSDeltaA} we show the behavior of the difference between the final and the initial horizon areas. The area always increases, as expected from the area law, which holds also in our model since the null energy condition is satisfied.

\section{Conclusion}
%
We have performed extensive nonlinear numerical simulations of the spherical collapse of (charged) wavepackets in flat spacetime and onto a charged black hole within Einstein-Maxwell theory and in an extension of the latter featuring nonminimal couplings and a spontaneous scalarization mechanism. First, within Einstein-Maxwell theory, we extended some previous analyses, confirming that no naked singularities form in these simulations and the final BH is always subextremal, in agreement with the cosmic censorship conjecture. We then extended this result to theories with spontaneous scalarization: although in that case it is possible to form scalarized BHs with charge above the RN bound, no naked singularities have been produced in all our simulations. 
A crucial role to prevent the formation of naked singularities and preserve the cosmic censorship is played by the (fully nonlinear) superradiance extraction of the BH charge and mass, which decreases the final BH charge-to-mass ratio.

Furthermore, we showed that hairy BHs can descalarize either by absorbing an opposite-charged wavepacket or by superradiant charge extraction, forming a subextremal RN BH. Overall, our results suggest that the cosmic censorship is at play also in Einstein-Maxwell-scalar theory featuring spontaneous scalarization.  As a by-product of our simulations, we also studied, at the full nonlinear level, the superradiant amplification of low-frequency charged wavepackets scattered off a charged (scalarized or not) BH.
In particular, the novel superradiantly-induced descalarization mechanism unveiled here deserves further studies. It would be interesting to explore whether it is at play in BH binaries to descalarize spin-induced scalarized BHs~\cite{Dima:2020yac,Berti:2020kgk,Herdeiro:2020wei} in modified gravity (see Ref.~\cite{Silva:2020omi} for dynamical descalarization in the context of BH binaries beyond GR), in which case it could have relevant astrophysical applications.

We expect that at least some of the phenomenology unveiled here for nonminimal Einstein-Maxwell-scalar theory would be similar for modified theories of gravity featuring the same scalarization mechanism, such as Einstein-scalar-Gauss-Bonnet gravity~\cite{Silva:2017uqg,Doneva:2017bvd,Antoniou:2017acq}. Recent advances in numerical simulations within these theories~\cite{East:2020hgw,East:2021bqk} can be used to perform similar gedankenexperiments as those presented here, thus challenging the cosmic censorship at the nonlinear level also in extensions of GR.

\begin{acknowledgments}
We acknowledge financial support provided under the European Union's H2020 ERC, Starting 
Grant agreement no.~DarkGRA--757480. We also acknowledge support under the MIUR PRIN and FARE programmes (GW-NEXT, CUP:~B84I20000100001), and 
from the Amaldi Research Center funded by the MIUR program ``Dipartimento di Eccellenza'' (CUP: 
B81I18001170001).
\end{acknowledgments}

\appendix

\section{Null Energy Condition} \label{app:NEC}
In this appendix we show that in Einstein-Maxwell-scalar theory with a positive coupling function and in spherical symmetry, the null energy condition is always satisfied. To prove this statement we have to show that
\begin{equation}
	T_{\mu\nu} m^\mu m^\nu \ge 0
	\label{eq:NEC}
\end{equation}
for any null vector $m^\mu$, where $T_{\mu\nu}$ is the total energy-stress tensor. The latter is made of three terms, respectively due to the real scalar field $\phi$, the complex scalar field $\xi$, and the electromagnetic field $F^{\mu\nu}$.

Let us consider these three terms separately. For two scalar fields one can show that
\begin{align}
	T^\text{\tiny SF}_{\mu\nu} m^\mu m^\nu &=  \frac{1}{4\pi} \bigl(m^\mu \nabla_\mu \phi \bigr)^2 \ge 0 \,, \\
	T^{\xi}_{\mu\nu} m^\mu m^\nu & = \frac{1}{2\pi} \bigl|m^\mu \D_\mu \xi \bigr|^2 \ge 0\,.
\end{align}
Finally, for the electromagnetic component we have
\begin{align}
	T^\text{\tiny EM}_{\mu\nu} m^\mu m^\nu &=  - \frac{1}{4\pi} m^\mu F_{\mu\alpha} \tensor{F}{^\alpha_\nu} m^\nu F[\phi].
\end{align}
Now, in spherical symmetry the magnetic field is absent and we can perform a 3+1 decomposition of the electromagnetic tensor as $F_{\mu\nu} = n_\mu E_\nu - n_\nu E_\mu$, where the electric field $E^\mu$ is orthogonal to $n^\mu$ (see Ref.~\cite{Alcubierre:2009ij}); therefore
\begin{align}
	T^\text{\tiny EM}_{\mu\nu} m^\mu m^\nu &= -\frac{1}{4\pi} F[\phi] m^\mu m^\nu \bigl[E_\mu E_\nu - n_\mu n_\nu (E_j E^j)\bigr]  \notag \\
				    &= \frac{1}{4\pi} F[\phi]\bigl[(m_\mu n^\mu)^2 (E_j E^j) - (m^i E_i)^2 \bigr],
	\label{eq:NECTEM}
\end{align}
Since $m^\mu$ is a null vector
\begin{align}
	0 = m_\mu m^\mu &= m_\mu \tensor{g}{^\mu_\nu} m^\nu = m_\mu (\tensor{\gamma}{^\mu_\nu} - n^\mu n_\nu) m^\nu = \notag \\
			&= m_i m^i - (m_\mu n^\nu)^2,
\end{align}
and thus $(m_\mu n^\nu)^2 = m_i m^i$. Substituting in Eq.~\eqref{eq:NECTEM} we obtain that if $F[\phi] \ge 0$ then
\begin{align}
	T^\text{\tiny EM}_{\mu\nu} m^\mu m^\nu &= \frac{1}{4\pi} F[\phi]\bigl[ m_i m^i (E_j E^j) - (m_i E^i)^2 \bigr] \ge 0,
\end{align}
where in the last step we used the Cauchy-Schwarz inequality. Since the term $T_{\mu\nu} m^\mu m^\nu$ can be decomposed in a sum of three positive terms then the null energy condition~\eqref{eq:NEC} is satisfied.

\section{Implementation of the PIRK integration scheme}\label{app:PIRK}
Here, we summarize the PRIK integration scheme. The equations of motion are written as~\cite{Montero:2012yr, Cordero-Carrion:2012qac}
\begin{equation}
	\begin{cases}
		\tder u &= \LL_1 (u, v) \\
		\tder v &= \LL_2 (u) + \LL_3(u, v) \\
	\end{cases},
\end{equation}
where $u$ schematically denotes the variables that are evolved fully explicitly whereas $v$ the variables that are evolved partially implicitly.

We used an analogous procedure to Ref.~\cite{Sanchis-Gual:2016tcm}. Namely we first evolved explicitly the variables $X$, $a$, $b$, $\alpha$, $\beta^r$, $E^r$, $\xi$ and $\phi$. As a second step we evolved partially implicitly $A_{a}$ and $K$, using
\begin{align}
	&\begin{cases}
		\LL_{2(K)} &= - D^2 \alpha, \\
		\LL_{3(K)} &= \beta^r \rder K + \alpha \Bigl( A_a^2 + 2 A_b^2 + \frac{1}{3} K^2 \Bigr) + \\
			&+ 4 \pi \alpha (S_a + 2 S_b + \mathcal{E}) \label{eq:Kb} \\ 
	\end{cases}, \\
	&\begin{cases}
		\LL_{2(A_a)} &=  - \Bigl(D^r D_r \alpha - \frac{1}{3} D^2 \alpha \Bigr) + \alpha (\tensor{R}{^r_r} - \frac{1}{3} R)  \\
		\LL_{3(A_a)} &=  \beta^r \rder A_a + \alpha K A_a - \frac{16 \pi \alpha}{3} \Bigl(S_a - S_b \Bigr) \\
	\end{cases},
\end{align}
Then, we evolved $\hat \Delta^r$, $\Pi$, $P$, $a_r$ and $\varphi$ using
\begin{align}
	&\begin{cases}
		\LL_{2(\hat \Delta^r)} &= \frac{2}{b} \rder \Bigl(\frac{\beta^r}{r} \Bigr) - 2\alpha \bigl( A_a - A_b \bigr) \frac{2}{br} + \\
				       &- \frac{2}{a} \bigl( A_a \rder \alpha + \alpha \rder A_a \bigr) + \\
				       &+ \frac{2 \alpha}{a} \Bigl[ \rder A_a + \bigl(A_a - A_b \bigr)\Bigl(\frac{\rder b}{b} + \frac{2}{r} \Bigr) + \\
				       &- 3 A_a \frac{\rder X}{X} - \frac{2}{3} \rder K \Bigl] + \\
				       &+ \frac{1}{a} \partial^2_r \beta^r + \frac{\sigma}{3} \frac{1}{a} \rder \hat \nabla_m \beta^m  \\
		\LL_{3(\hat \Delta^r)} &= \beta^r \rder \hat \Delta^r - \hat \Delta^r \rder \beta^r + 2 \alpha A_a \hat \Delta^r + \\
				       &+ 2 \frac{\sigma}{3} \hat \Delta^r \hat \nabla_m \beta^m - \frac{16 \pi \alpha}{a} j_r\\
	\end{cases}, \\
	&\begin{cases}
		\LL_{2(\Pi)} &= \frac{\alpha X^2}{a} \Bigl[ \bigl(\partial_r \phi\bigr)\Bigl(\frac{2}{r} - \frac{\partial_r a}{2a} + \frac{\partial_r b}{b} - \frac{\rder X}{X}\Bigr) + \\
		&+ \partial_r^2 \phi \Bigr] + \frac{(\partial_r \phi)(\partial_r \alpha)}{a} X^2 + \frac{1}{2} \frac{\alpha a}{X^2} (E^r)^2 \frac{\delta F[\phi]}{\delta \phi} \\
		\LL_{3(\Pi)} &= \beta^r \partial_r \Pi + \alpha \Pi K \\
	\end{cases}, \\
	&\begin{cases}
		\LL_{2(P)} &= \frac{\alpha X^2}{a} \Bigl[ \bigl(\partial_r \xi\bigr)\Bigl(\frac{2}{r} - \frac{\partial_r a}{2a} + \frac{\partial_r b}{b} - \frac{\rder X}{X} \Bigr) + \\
			   &+ \partial_r^2 \xi \Bigr] + \frac{(\partial_r \xi)(\partial_r \alpha)}{a} X^2 + 2 i q \alpha \Bigl( \varphi P + \frac{a_r \rder \xi}{a} X^2 \Bigr) + \\
			   & - q^2 \alpha \Bigl( \frac{(a_r)^2}{a} X^2 - \varphi^2 \Bigr) \xi \\
		\LL_{3(P)} &=\beta^r \partial_r P + \alpha P K
	\end{cases}, \\
	&\begin{cases}
		\LL_{2(a_r)} &= a_r \rder \beta^r - \rder (\alpha \varphi) - \frac{\alpha a}{X^2} E^r \\
		\LL_{3(a_r)} &= \beta^r \rder a_r
	\end{cases}, \\
	&\begin{cases}
		\LL_{2(\varphi)} &= - \frac{\alpha X^2}{a} \Bigl[ a_r \Bigl(\frac{2}{r} - \frac{\rder a}{2 a} + \frac{\rder b}{b} - \frac{\rder X}{X} \Bigl) + \\
				 &+ \rder a_r \Bigr] - \frac{(\rder \alpha ) a_r}{a} X^2 \\
		\LL_{3(\varphi)} &=  \beta^r \rder \varphi + \alpha \varphi K \\
	\end{cases}.
\end{align}
Finally, we evolved $B^r$ fully implicitly.

\section{Convergence tests}\label{app:convergence}
We checked the convergence of our code by computing the violation of the Hamiltonian constraint and studying its scaling with respect to the grid step. 

We evolved the evolution equations using the initial condition discussed in Sec.~\ref{sec:scalarization}, setting the initial BH charge-to-mass ratio to $0.9$, and $q M = 5$. The grid extends from the origin up to $r_\infty = 250 M$, and the grid steps we used are $\Delta r = 0.01 M$ and $\Delta r = 0.005 M$. In both cases the CFL factor was set to ${\rm CFL} = 0.4$. 

\begin{figure}
	\centering
	\includegraphics[width = \columnwidth]{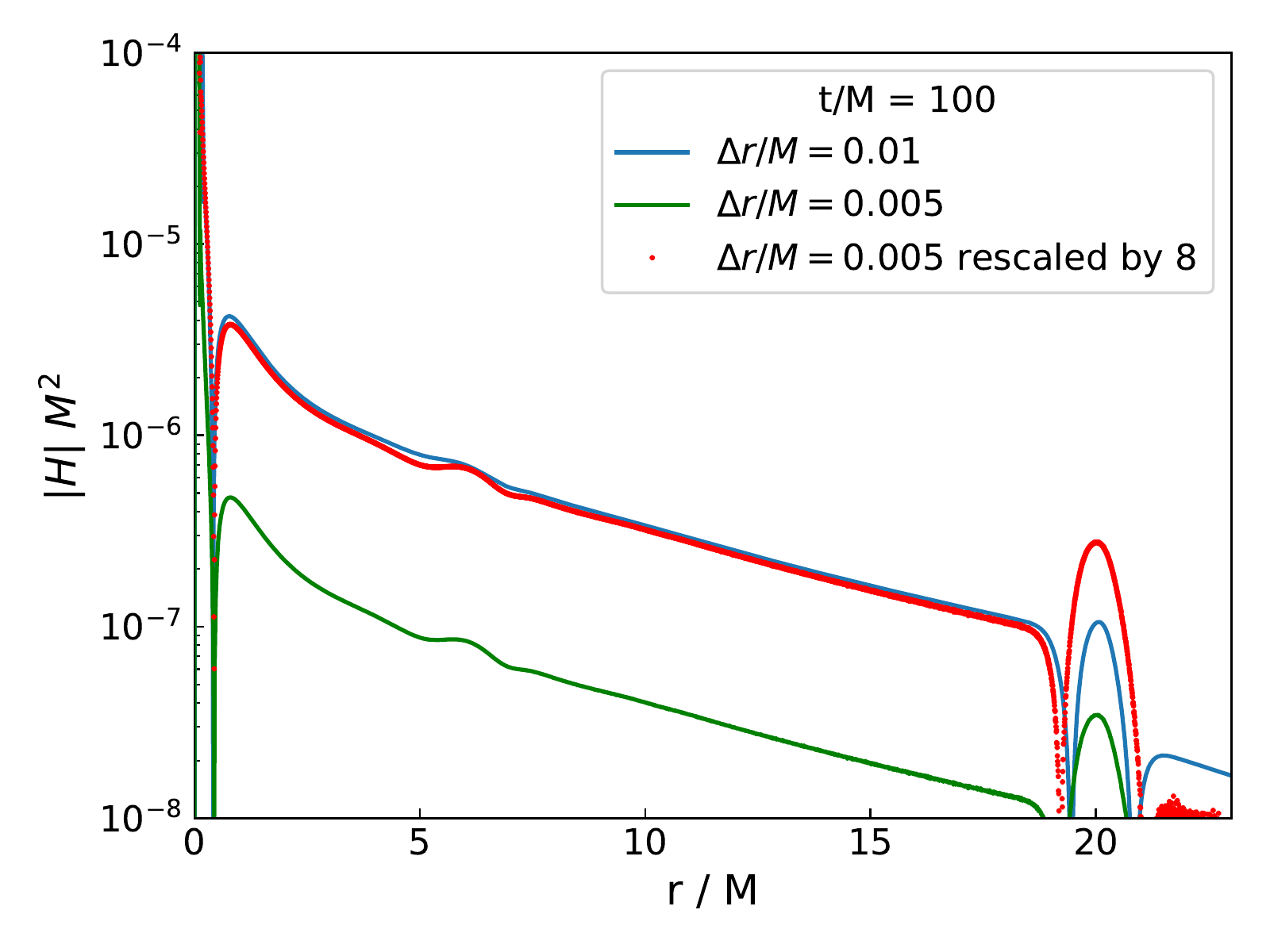} \\
	\includegraphics[width = \columnwidth]{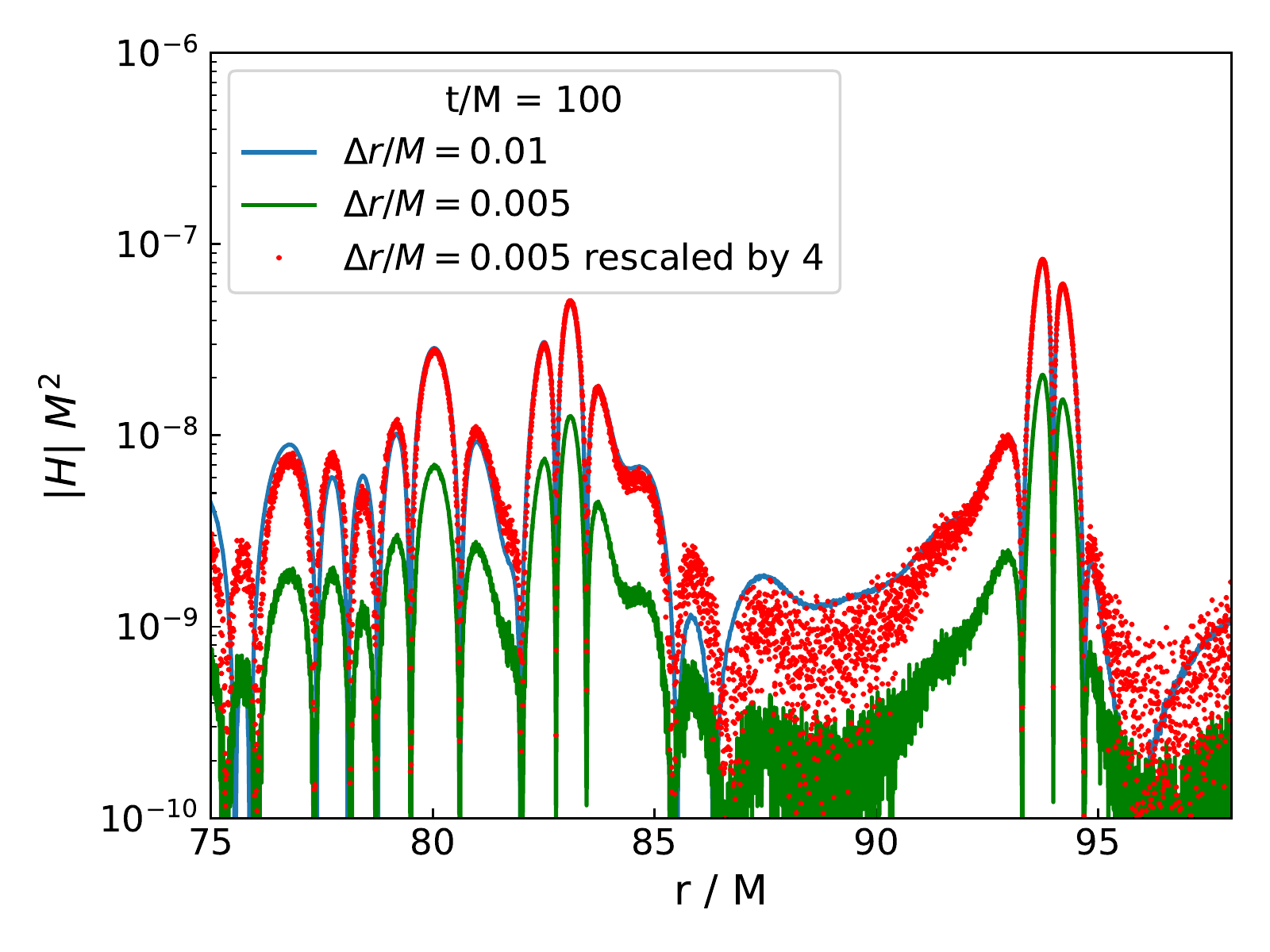}
	\caption{Convergence of the code. Continuous lines denote the violation of the Hamiltonian constraint for the two spatial resolutions, while the dots denote the behavior corresponding to the higher resolution rescaled by the factor indicated in the legend. These plots show third-order convergence near the horizon, and second-order convergence in the outer region.}
	\label{fig:convergence}
\end{figure}

We then computed the violation of the Hamiltonian constraint at $T = 100 M$, and the results are shown in Fig.~\ref{fig:convergence}. As we can see from the plot, near the horizon the violation of the Hamiltonian constraint behaves as a third-order term, while in an outer region it scales as a second-order term, in agreement with the order of our numerical scheme.

\bibliographystyle{apsrev4}
\bibliography{References}

\end{document}